\documentclass[a4paper,12pt]{article}

\usepackage[cp1251]{inputenc}
\usepackage[T1,T2A]{fontenc}
\usepackage[english,russian]{babel}

\usepackage{fullpage}
\usepackage{amsmath,amssymb,mathrsfs}
\usepackage{graphicx}

\begin{document}

\title{Необычный магнетизм решеток Кондо}

\author{В. Ю. Ирхин}

\maketitle

\begin{abstract}
Дан обзор экспериментальной ситуации и теоретических представлений относительно необычного магнитного упорядочения плотных кондовских $4f$- и $5f$-систем, включая недавно исследованные тройные системы на основе церия, иттербия и актинидов. Оно характеризуется малым магнитным моментом и проявляет черты, характерные как для магнетиков с локализованными спинами, так и для систем с коллективизированными электронами. Особое внимание уделено ферромагнитным системам, а также конкуренции различных типов магнитного упорядочения. Рассмотрены методы теории возмущений и ренормгруппы, а также представление вспомогательных псевдофермионов для описания формирования основного состояния решеток Кондо. Обсуждается проблема нефермижидкостного поведения.
\end{abstract}

Содержание

1. Введение

2. Обсуждение экспериментальных данных

2.1 Цериевые системы

2.2 Актинидные системы

2.3 Иттербиевые системы

3. Модели, теория возмущений и ренормгруппа для решеток Кондо

4. Магнетизм и нефермижидкостное поведение. Многоканальная модель Кондо

5. Проблема основного состояния. Приближение среднего поля в представлении псевдофермионов

6. Заключение

Список литературы

\section{Введение}

Системы с тяжелыми фермионами (гигантскими значениями электронной теплоемкости) и другие $4f$- и $5f$-соединения с необычными свойствами  остаются интереснейшими объектами исследований в многоэлектронной физике твердого тела: они представляют собой весьма нетривиальный пример сильно коррелированных систем, где реализуются экзотические состояния вещества. С точки зрения теории их обычно рассматривают как решетки Кондо или плотные (концентрированные) кондовские системы, т.е. периодические решетки $f$-спинов, где взаимодействие с электронами проводимости приводит к экранированию и подавлению локализованных магнитных моментов и аномалиям электронных свойств \cite{Brandt} (инфракрасная катастрофа, впервые обнаруженная Кондо~\cite{Kondo}). Для теоретического описания этих систем используется весь арсенал современных методов квантвой теории поля.

С другой стороны, экспериментальные исследования показывают, что среди таких соединений широко распространено магнитное упорядочение, часто с малым магнитным моментом, и/или развитые спиновые флуктуации.
Для систем с "<умеренно"> тяжелыми фермионами (коэффициент в линейном члене теплоемкости $\gamma$ порядка 100 мДж/моль К$^2$) достаточно типично сосуществование неполного кондовского экранирования с аномальным магнитным упорядочением (с резко подавленным, но ненулевым моментом).
Особенно часто встречается антиферромагнитное упорядочение, хотя нередко и ферромагнитное. Библиографию и обсуждение ранних экспериментальных данных по конкретным соединениям и соответствующих теоретических представлений можно найти в книге \cite{II} и статье \cite{601}.

Класс "<кондовских"> магнетиков характеризуют следующие особенности~\cite{601,IK}:
\begin{enumerate}
    \item Логарифмическая температурная зависимость удельного
сопротивления при $T>T_{K}$, присущая кондо-системам. Эта зависимость может быть получена в третьем порядке теории возмущений, как и в оригинальном подходе Кондо \cite{Kondo}.
    \item Малое значение магнитной энтропии в точке
упорядочения по сравнению со значением $R\ln{} (2J+1)$ ($R$ --- универсальная газовая постоянная, $J$ --- номинальное значение магнитного момента), которое
соответствует обычным магнетикам с локализованными моментами. Это
явление связано с подавлением магнитной теплоемкости вследствие
эффекта Кондо (экранирования моментов): лишь малая часть изменения
энтропии связана с дальним магнитным порядком.
    \item Упорядоченный магнитный момент $\mu_s$ мал по
сравнению с высокотемпературным моментом $\mu _{\text{eff}}$,
найденным из постоянной Кюри. Последний имеет, как правило,
нормальное значение, близкое к соответствующему значению для
редкоземельного иона (например $\mu _{\text{eff}}\simeq 2{,}5\mu
_{\text{B}}$ для иона Ce${}^{3+}$). Такое поведение напоминает
слабые коллективизированные (зонные) магнетики.
    \item Парамагнитная точка Кюри $\theta $, как правило,
отрицательна (даже для ферромагнетиков) и заметно превышает по
абсолютной величине температуру магнитного упорядочения. Такое поведение может быть обязано большому одноузельному кондовскому вкладу в парамагнитную
восприимчивость $\chi$, который доминирует над межузельными обменными взаимодействиями. В частности, интерполяционная формула из точного решения модели Кондо, справедливая при $0.5T_{K}<T<16 T_{K}$, имеет вид $\chi (T)=0.17/(T+1.5 T_{K})$, где $T_{K}$~"--- температура Кондо по Вильсону \cite{558}. С другой стороны, определенную роль могут играть фрустрации обменных взаимодействий.
\end{enumerate}

Что касается "<классических"> систем с тяжелыми фермионами (с большим $\gamma > 400$ мДж/моль К$^2$ \cite{Stewart}), здесь ситуация сложнее. Существуют ясные свидетельства антиферромагнетизма в UCd${}_{11}$ и U${}_2$Zn${}_{17}$ с тем~же
порядком величины $\mu_s$~\cite{Stewart}. Для соединений
UPt${}_3$ и URu${}_2$Si${}_2$ моменты насыщения крайне малы: $\mu_s\simeq
2-3\cdot 10^{-2}\mu _{\text{B}}$. Признаки антиферромагнитного
упорядочения с очень малым $\mu_s$ также наблюдалось для
CeAl${}_3$, UBe${}_{13}$, CeCu${}_2$Si${}_2$, CeCu${}_6$ (впрочем,
ряд данных для этих систем подвергались сомнению, см. также обзор
\cite{Vojta1}).

Типичная особенность тяжелофермионных магнетиков~"---
высокая чувствительность $\mu_s$ к внешним параметрам,
таким, как давление и легирование малым количеством примесей.
Например, UBe${}_{13}$ становится антиферромагнитным с заметным
$\mu_s$ под давлением $P>23$~кбар; напротив, CeAl${}_3$
становится парамагнитным под давлением выше $P=3$~кбар. Момент в
UPt${}_3$ увеличивается до значений порядка одного $\mu
_{\text{B}}$ при замене~$5\,\%$ Pt на Pd или $5\,\%$ U на Th. Ряд
систем с тяжелыми фермионами претерпевают метамагнитные переходы в
слабых магнитных полях с резким увеличением магнитного момента.
Здесь характерно название обзора П. Коулмена
\cite{Coleman1}: "<Тяжелые фермионы. Электроны на грани
магнетизма">.

В 1990-е годы был открыт еще одни класс $f$"~-электронных систем:
для ряда соединений и сплавов на основе церия и урана было
обнаружено нефермижидкостное (НФЖ) поведение~\cite{646,646a,NFL}. Оно
проявляется в необычных температурных зависимостях электронной теплоемкости $C(T)$ вида $T \ln T$ или $T^{1-\lambda}$ с малыми показателями  $\lambda$, аномальном степенном поведении магнитной восприимчивости $ T^\zeta $ ($\zeta<1$) и
сопротивления $T^{\mu}$ ($\mu<2$), и~т.~д.
Часто НФЖ поведение возникает на границе магнитного упорядочения (квантового фазового перехода) \cite{Vojta1}, хотя обсуждаются и его другие разнообразные механизмы.

В данном обзоре рассмотрены современные теоретические представления и экспериментальные данные по необычному магнитному упорядочению плотных кондовских $f$-систем и его проявлению в электронных свойствах, включая нефермижидкостное поведение.

\section{Экспериментальные результаты}

Имеются многочисленные примеры соединений, где кондовские аномалии
в термодинамических и кинетических свойствах сосуществуют с
магнитным упорядочением, а момент насыщения $\mu_s$ имеет
величину порядка магнетона Бора. Это ферромагнетики CePdSb,
CeSi${}_x$, Sm${}_3$Sb${}_4$, Ce${}_4$Bi${}_3$, NpAl${}_2$,
антиферромагнетики CeAl${}_2$, TmS, CeB${}_6$, UAgCu${}_4$ (см. более подробное обсуждение в \cite{II}); встречаются и системы с более экзотическими магнитными свойствами.  Ниже будут приведены последние результаты по магнетизму кондовских систем.


\subsection{Цериевые системы}

После активного исследования классических систем с тяжелыми фермионами и решеток Кондо, в 1980-е годы настало время для рассмотрения тройных систем.
Один из наиболее ярких примеров здесь~"--- кондовский ферромагнетик CeRh$_3$B${}_2$
с~c температурой Кюри $T_C=115$~К, парамагнитной температурой Кюри $\theta =-370$~К~\cite{284} и сравнительно небольшим
$\gamma =16$~мДж/моль${}\cdot {}$К${}^2$.

Постоянно обнаруживаются новые антиферромагнитные системы с низкой температурой Нееля и редуцированным моментом основного состояния,
например Ce$_{8}$Pd$_{24}$Ga ($T_N=$ 3.6 K, $\mu_s = 0.36 \mu_B$/ат. Ce \cite{Ce8N24Ga}),  слоистые системы Ce$_3$(Pd,Pt)In$_{11}$ \cite{Ce3PdIn11}.

Растет и число кондовских феррмагнетиков:
CePt \cite{CePt}, CeRu$_2$Ge$_2$ \cite{CeRuGe}, CeAgSb$_2$, \cite{CeAgSb},  CeRuPO \cite{CeRuPO}, CeRu$_2$M$_2$X (M = Al, Ga; X = B, C) \cite{CeRu2Ga2B1,CeRu2Ga2B}, CeIr$_2$B$_2$ \cite{CeIr2B2}, гидрогенированный CeNiSn \cite{CeNiSn},
CeFe$_4$Sb$_{12}$ \cite{CeFe4Sb12},
Ce$_{4}$Sb$_{1.5}$Ge$_{1.5}$ ($T_C = 13$~K, $\theta = -9$~K) \cite{Ce4Sb3}.

В системе Ce$_{1-x}$La$_{x}$PdSb, где имеется высокотемпературный логарифмический вклад в сопротивление, ферромагнитное кондо-состояние постепенно переходит в НФЖ в области $x>0.7$, причем последнее по-видимому носит одноузельный характер \cite{CePdSb}.


Хороший пример конкуренции эффекта Кондо, антиферро~"--- и ферромагнетизма дает система (Ce$_{1-x}$Nd$_{x}$)$_3$Al \cite{(Ce1-xNdx)3Al}. Здесь при $x>0.2$ антиферромагнитное упорядочением сменяется ферромагнитным, причем при низких температурах эффект Кондо и ферромагнетизм сосуществуют при $0.2<x<0.3$.

В работах \cite{CeRu2X2M1,CeRu2X2M2,CeRu2X2M3} был исследован ряд соединений CeRu$_2$X$_2$M (X=Al, Ga и M=B, C), которые демонстрируют ферромагнетизм с $T_C $ от 12.8 K (CeRu$_2$Al$_2$B) до 17.2 K (CeRu$_2$Ga$_2$C).
Электронная теплоемкость этих систем невелика ($\gamma$ около 30 мДж/моль K$^2$), так что проявления эффекта Кондо являются слабыми.
Согласно \cite{CeRu2Al2B}, в системе CeRu$_2$A$_2$B при $T_C <T<T_N=14.2K$  имеется несоизмеримый магнитный порядок.
Последний возникает и в ряде других кондовских цериевых систем, например, в CeAu$_2$Ge$_2$ \cite{CeAu2Ge2}, Ce$_2$M$_m$In$_{3n+2m}$ \cite{CenMmIn3n+2m}.

В системе CeCd$_{0.7}$Sb$_{2}$ \cite{CeCd07Sb2} наблюдается сильно анизотропный ферромагнетизм с $T_C = 3$K и отрицательной парамагнитной температурой Кюри $\theta = -24$K, причем при $T_N = 0.8$~K происходит переход в антиферромагнитное состояние. Аналогичная ситуация имеет место в системе CeZn$_{0.66}$Sb$_{2}$ \cite{CeZn066Sb2} ($T_C = 3.6$~K, $T_N = 0.8$~K, $\theta = -11$~K).

Тройные интерметаллические соединения на основе церия типа  CeTX$_2$ (T – переходный металл, X = Si, Ge, Sn) интересны благодаря необычным свойствам их основного состояния. В частности, CePtSi$_2$  – система с тяжелыми фермионами, где значение коэффициента Зоммерфельда достигает $\gamma=$ 1.7 Дж/моль K$^2$ при 1.25 K \cite{CePtSi}, CeRhSi$_2$ \cite{CeRhSi} и CeNiSi$_2$ \cite{CeNiSi} -- системы с сильными валентными флуктуациями. Описание таких интерметаллидов должны строиться на основе учета ряда факторов: окружения церия, гибридизации 4f-уровня с зоной проводимости, величины констант обменного взаимодействия.

Система CeRuSi$_2$ была подробно исследована в работах  \cite{Physica,Muon,CeRuSi2,CeRuSi21}.
Сосуществование эффекта Кондо и ферромагнетизма с малыми  моментами  порядка 0.1$\mu_B$, возникающего ниже 11.2 К, подтверждается как исследованием кинетических свойств \cite{CeRuSi2}, так и мюонными измерениями \cite{Muon}).
Полевые зависимости намагниченности вплоть до очень сильных полей 150 Т показали отсутствие насыщения намагниченности, что напоминает поведение слабых коллективизированных магнетиков.
\begin{figure}[htbp]
\includegraphics[width=3.3in, angle=0]{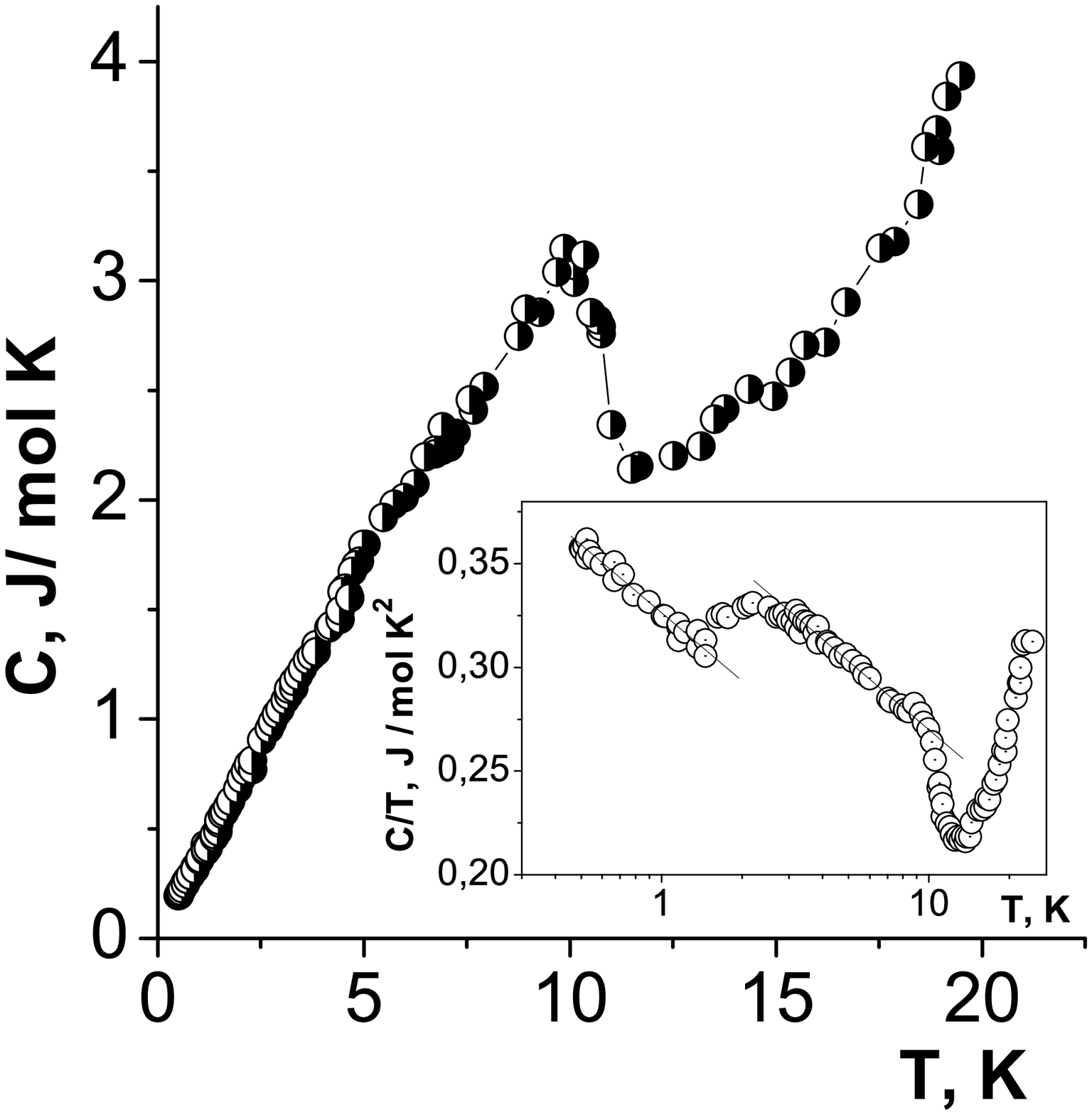}
\caption{Поведение теплоемкости CeRuSi$_2$.
На вставке -- теплоемкость в логарифмическом масштабе}
\label{112Cmag}
\end{figure}
На рис. \ref{112Cmag} показано температурное изменение теплоемкости CeRuSi$_2$~\cite{CeRuSi21}. На фоне достаточно высокой электронной теплоемкости магнитная теплоемкость  имеет $\lambda$-пик в районе 11 К (соответствующее изменение энтропии $\Delta \mathcal{S} = 2.7$~Дж/ моль K  = 0.5 $R\ln 2$). Значение коэффициента Зоммерфельда, полученное путем стандартной экстраполяции зависимости $C/T$ (выше температуры 20 К), составляет $\gamma \simeq$ 140 мДж/моль K$^2$, что соответствует умеренно тяжелым фермионам. При низких температурах $\gamma$ растет и имеются указания на логарифмическое нефермижидкостное поведение теплоемкости, что особенно интересно для стехиометрического соединения (аномалия вблизи $T = 2$~K может быть связана с влиянием кристаллического поля).

В работе \cite{CeRuSi21}  был также исследован большой набор соединений CeRuX$_2$, CeRu$_2$X$_2$, Ce$_2$Ru$_3$X$_5$ и CeRuX$_3$  (X = Si, Ge).
В системе CeRu$_2$Si$_2$ не наблюдается магнитного упорядочения, но есть кондовский максимум сопротивления при температуре 50 K (в CeRuSi$_2$ он гораздо выше – при 150 K). Это различие вероятно связано с более низкой температурой Кондо (объем элементарной ячейки в CeRu$_2$Si$_2$ существенно больше).
Следует отметить, что метамагнетизм, ранее обсуждавшийся для CeRu$_2$Si$_2$,
теперь рассматривается  как непрерывный кросссовер \cite{Daou,Matsuda}, причем коллективизированный характер 4f-электронов сохраняется до очень сильных полей около 50 T \cite{Matsuda}, что напоминает ситуацию в CeRuSi$_2$.

\subsection{Иттербиевые системы}

Особенно часто ферромагнитное упорядочение в сочетании с проявлениями эффекта Кондо встречается в тройных системах на основе иттербия.
Иногда оно связано с формированием  неколлинеарной магнитной структуры – например, в YbNiSn возникает малый скошенный ферромагнитный момент \cite{YbNiSn}.

В ряде аномальных иттербиевых систем наблюдается линейное по температуре сопротивление, свидетельствующее о нефермиждкостном поведении.
Интересно, что подобное аномальное поведение сопротивления  наблюдается и в системе YbMn$_2$Sb$_2$ с немагнитным иттербием, где обычный эффект Кондо должен отсутствовать. Эти аномалии могут быть связаны с рассеянием на псевдоспиновых степенях свободы \cite{NikiforovYb}.

В работе \cite{NikiforovFTT} были получены и исследованы новые тройные соединения на основе иттербия: YbPd$_2$Ge, YbPd$_2$Si, YbPdGe, YbPdSi, YbPtGe, а также YbPdGe$_2$ и ранее известное соединение YbPd$_2$Si$_2$. Из гальваномагнитных свойств и магнитной восприимчивости было обнаружено, что германиевые соединения YbPdGe, YbPtGe и YbPd$_2$Ge при низких температурах (вблизи 16, 10, и 12 K соответственно) демонстрируют ферромагнитное упорядочение.

В более поздних работах результаты по магнитным и термодинамическим свойствам были дополнены. В работе \cite{YbPtGe} для YbPtGe были найдены значения $T_C = 5.4$K, $\gamma = $209 мДж/моль K$^2$ ниже $T_C $. Магнитная энтропия в $T_C $ составляла 0.52 $R\ln 2$. Для YbPdGe в \cite{YbPdGe1} были получены значения $T_C$  = 11.4 K и $\gamma$ = 150 мДж/моль K$^2$; ситуация в этой системе напоминает CeRuSi$_2$.

В работе \cite{YbPdGe} обсуждаются различные проявления Кондо в YbPdGe и YbPtGe, включая логарифмический вклад в сопротивление, и проводится сравнение кинетических свойств с соответствующими цериевыми системами CePdGe и CePtGe (формально трехвалентные ионы церия и иттербия  схожи, поскольку соответствуют одному электрону или дырке в 4f-оболочке).
Как показано в \cite{CeTX}, соединения CeTX (T = Pd и Pt, X = Ga, Ge и Sn) являются кондовскими системами, антиферромагнитными при низких температурах -- в противоположность YbTGe.
В обзоре \cite{Flouquet} различие этих случаев подробно обсуждается на примере YbRh$_2$Si$_2$  -- типичного антиферромагнитного  соединения иттербия с тяжелыми фермионами. Несмотря на сходство с CeRh$_2$Si$_2$ (близкие значения температуры Кондо $T_K$ и амплитуды максимума сопротивления, линейность сопротивления выше $T_N$), имеются и заметные различия, которые можно связать с большей локализацией (меньшей степени гибридизации с электронами проводимости), а также более сильным спин-орбитальным взаимодействием в случае 4f-орбиталей иттербия.

Согласно работе \cite{YbPdSi}, система YbPdSi является тяжелофермионным ферромагнетиком с $T_C = 8$~K и $\gamma$  около 300 мДж/моль K$^2$.
Исследование соединений YbTGe (T=Rh, Cu, Ag) \cite{YbTGe} показало, что YbRhGe является антиферромагнетиком с $T_N=7$~K, YbCuGe – ферромагнетиком с моментом $0.7\mu_B$ и $T_C=8$~K; в YbAgGe было найдено очень большое $\gamma = $570 мДж/моль K$^2$. Впоследствии в YbAgGe было обнаружено низкотемпературное неколлинеарное магнитное упорядочение \cite{YbAgGe}.
Согласно \cite{YbAgGe1}, YbAgGe  -- фрустрированный тяжелофермионный антиферромагнетик со сложной магнитной фазовой диаграммой, проявляющий НФЖ поведение.
В антиферромагнитной системе Yb(Rh$_{1-x}$Co$_x$)$_2$Si$_2$  был обнаружен ферромагнетизм при $x=0.27$ ниже  $T_C=1.3$~K с моментом $0.1 \mu_B$ \cite{DopedYbRh2Si2}.


Системы Yb$_2$(Pd$_{1-x}$Ni$_x$)$_2$Sn \cite{Yb2(Pd1?xNix)_2Sn} и Yb$_2$Pd$_2$In$_{1-x}$Sn$_x$ \cite{Yb2Pd2In1-xSnx} демонстрируют фрустрированный магнетизм – необычное магнитное упорядочение возникает при легировании. В последней системе упорядочение возникает в узкой области вблизи $ x = 0.6$ на фоне нефермижидкостного поведения.

В квазиодномерной системе YbNi$_4$P$_2$  был обнаружен экзотический ферромагнетизм с $T_C=0.17$ K и крайне малым моментом насыщения 0.05 $\mu_B$ \cite{YbNi4P2}. При этом сопротивление линейно выше $T_C$, а коэффициент $\gamma$ расходится сильнее, чем логарифмически.

\subsection{Актинидные системы}

Металлы группа урана проявляют сильную тенденцию к паулиевскому парамагнетизму в силу своей электронной структуры. Однако их магнетизм может существенно усиливаться в соединениях с переходными металлами: как благодаря увеличению межатомного расстояния (согласно критерию Хилла,  это способствует формированию локализованных моментов), так и увеличению плотности состояний вблизи уровня Ферми \cite{371}.
Важным отличием аномальных урановых соединений от цериевых является более сильная делокализация 5f-электронов, что отражается и на спектре спиновых флуктуаций, которые становятся более сильными \cite{Lawrence}.

Сосуществование кондовских аномалий и дальнего магнитного порядка в соединениях актинидов не является редким.
В ряде актинидных систем ферромагнитное упорядочение сосуществует с кондовскими логарифмическими вкладами в сопротивление (см. также обсуждение в \cite{Perkins}).
Вот некоторые примеры: UTe ($T_C = 102$~K \cite{UTe}), UAsSe ($T_C = 109$~K \cite{UAsSe}), UCu$_{0.9}$Sb$_2$ ($T_C = 113$~K \cite{UCu0.9Sb2}),
UCo$_{0.5}$Sb$_2$ ($T_C =64.5$~K \cite{UCo0.5Sb2}),
UNiSi$_2$  ($T_C = 95$~K \cite{UNiSi2}),
NpNiSi$_2$ ($T_C =51.5$~K \cite{NpNiSi2}).
В соединении Np$_2$PtGa$_3$ ($T_C= 26$~K) наблюдается высокое значение $\gamma$ = 180 мДж/моль Np K$^2$ \cite{Np2PtGa3}.

Система UIr$_2$Zn$_{20}$ \cite{UIr2Zn20} по праву может быть названа ферромагнитной системой с тяжелыми фермионами. Здесь $\gamma $= 450 мДж/моль K$^2$ и остается высокой ниже точки Кюри $ T_C= 2.1 $ K, а момент насыщения составляет $0.4 \mu_B$.

В соединении UAu$_2$Si$_2$ ниже ($T_C =19$~K наблюдается нескомпенсированный антиферромагнетизм со спонтанной намагниченностью вдоль оси $c$, причем при низких температурах $\gamma $= 180 мДж/моль K$^2$ \cite{UAu2Si2}.

Ряд актинидных систем демонстрируют нефермидкостное поведение.
В системе Th$_{1-x}$U$_{x}$Cu$_{2}$Si$_{2}$ \cite{UCuSi2,NFL} НФЖ поведение ($C/T$ пропорционально $\ln T$, а сопротивление линейно по температуре) возникает в области составов, где ферромагнетизм подавлен.

Соединение URhGe ферромагнитно. В сплаве URh$_{1-x}$Ru$_x$Ge  \cite{URhGe} критическая концентрация подавления магнитного порядка равна $x_{cr}$ = 0.38, температура Кюри обращается в нуль по линейному закону с ростом $x$, а ферромагнитный момент
$\mu_s $ исчезает непрерывным образом ($\mu_s = 0.4 \mu_B$ для $x= 0$). В точке $x_{cr}$ теплоемкость ведет себя как $T \ln T$, а показатель   в степенной зависимости сопротивления от температуры достигает минимума $\mu=1.2$. Полная магнитная энтропия, полученная интегрированием  $C/T$, составляет $0.48R\ln2$ для $x=0$ и падает до $0.33R\ln2$ при $x_{cr}$.

В то же время соединение NpRhGe \cite{NpRhGe} является антиферромагнетиком с $T_N = 21$~K, упорядоченным моментом $\mu_{\rm Np}= 1.14 \mu_{B}$ и высоким значением $\gamma $ = 195 мДж/моль K$^2$.

В системе UCoGe ферромагнетизм также сменяется НФЖ состоянием при замене кобальта рутением \cite{UCoGe}.

Еще один пример нефермижидкостного поведения в ферромагнитной фазе дает тяжелофермионная система URu$_{2-x}$Re$_x$Si$_2$ \cite{URu}. В ней были обнаружены малый момент основного состояния $\mu_s =0.44$ $\mu_B$ и зависимости  $- \ln T$ (или $T^{-0.1}$) для $C/T$ и $T^\mu $ с $\mu =1.2$ для сопротивления в широком интервале температур  ниже 20 К для $x=$ 0.6. В то же время аномалий теплоемкости и сопротивления в точке магнитного перехода здесь не наблюдалось.

Получение чистых кристаллов UGe$_2$ и в последнее время из URhGe привело к открытию ферромагнитных сверхпроводников. Считается, что решающее значение для сосуществования ферромагнетизма и сверхпроводимости имеет триплетное спаривание \cite{Flouquet,URhGe,UFN}.

\section{Модели, теория возмущений и ренормгруппа для решеток Кондо}

Долгое время считалось, что конкуренция межузельного обменного
РККИ-взаимодействия и эффекта Кондо должна привести к формированию
или обычного магнитного упорядочения с большими атомными магнитными
моментами (как в чистых редкоземельных металлах), или немагнитного
состояния Кондо с подавленными магнитными моментами \cite{Brandt}. Это рассуждение основывалось на критерии Дониаха \cite{Don}, в котором сравниваются энергетические масштабы этих фаз.
Однако приведенные выше экспериментальные данные заставили пересмотреть эту точку зрения и привели к развитию более сложных теоретических подходов.

Рассмотрим гамильтониан $s-f$~обменной модели,

\begin{equation}
H=H_s +H_f+ H_{\text{int}}= \sum_{\mathbf{k}m\sigma}
t_{\mathbf{k}}c_{\mathbf{k}m\sigma }^{\dagger }c_{\mathbf{k}m\sigma}
-\sum_{\mathbf{q}}J_{\mathbf{q}}\mathbf{S}_{-\mathbf{q}}
\mathbf{S}_{\mathbf{q}}
-I\sum_{im\sigma \sigma ^{\prime}}(\mathbf{S}_i\mbox{\boldmath$\sigma $}_{\sigma \sigma ^{\prime }})
c_{im\sigma }^{\dagger }c_{im\sigma ^{\prime }},
 \label{eq:G.2}
\end{equation}
где $c_{\mathbf{k}m\sigma }^{\dagger }$, $c_{\mathbf{k}m\sigma }$ --  операторы электронов проводимости, $t_{\mathbf{k}}$~"--- затравочная зонная энергия,
$\mathbf{S}_i$ -- операторы локализованных спинов,
{\boldmath$\sigma $}~"--- матрицы Паули, $I$~"--- параметр $s$--$f$~обменного взаимодействия.
Часто (например, в~редкоземельных металлах) $f-f$ взаимодействие между локализованными спинами $J_{\mathbf{q}}$
является косвенным РККИ-обменом через электроны проводимости,
причиной которого служит то~же самое $s$---$f$~взаимодействие (хотя в ряде соединений может играть важную роль и сверхобменное взаимодействие).
Однако при построении теории возмущений удобно включить $f-f$ взаимодействие в нулевой гамильтониан (такая модель получила в зарубежной литературе название модели Кондо-Гейзенберга).

В отличие от внутриатомного кулоновского (хаббардовского)
взаимодействия, $s- f$ обменное взаимодействие как правило не является
сильным, однако приводит к существенным эффектам в
электронном спектре. С микроскопической точки зрения оно может
иметь различную природу. В ряде редкоземельных систем (например, в
магнитных полупроводниках) это внутриатомный хундовский обмен,
который ферромагнитен. В редкоземельных соединениях $s- f$ обмен часто является не
настоящим, а эффективным -- обусловленным гибиридизацией между
 $s$-зонами проводимости и атомными уровнями $ f$-электронов; в этом
случае он антиферромагнитен ($I<0$). Последнее является необходимым для возникновения эффекта Кондо: при этом условии эффективное (перенормированное) обменное взаимодействие становится
бесконечным в режиме сильной связи, так что магнитное рассеяние приводит к полному
экранированию магнитных моментов \cite{556,558}.

В гамильтониане (\ref{eq:G.2}) для общности учтено орбитальное вырождение: введен индекс $m=1…M$, где $M$ -- число каналов рассеяния электронов проводимости, которое может быть дополнительным формальным большим параметром при построении теории возмущений.
Как впервые показали Нозьер и Бландин при анализе расщепления уровней в кристаллическом поле в случае одной кондовской примеси \cite{555}, скейлинговое поведение существенно зависит от соотношения величины локализованного спина и орбитального вырождения электронов. В случае $S>M/2$ имеем ситуацию недокомпенсации (неполного экранирования), когда в режиме сильной связи $S \rightarrow S-M/2$. При $S=M/2$ имеем обычный эффекта Кондо (полное экранирование, формирование тяжелой ферми-жидкости), а случай $S<M/2$ (перекомпенсация) наиболее сложен – здесь возникает нетривиальная фиксированная точка и нефермижидкостное поведение уже в однопримесном режиме.

В более общей $SU(N)\otimes SU(M)$ модели мы имеем $\sigma =1...N$ и гамильтониан записывается как \cite{Cox}
\begin{equation}
H=\sum_{\mathbf{k}m\sigma }t_{\mathbf{k}}c_{\mathbf{k}m\sigma }^{\dagger }c_{%
\mathbf{k}m\sigma }^{{}}-I\sum_{im\sigma \sigma ^{\prime }}|i\sigma ^{\prime
}\rangle \langle i\sigma |c_{im\sigma }^{\dagger }c_{im\sigma ^{\prime
}}^{{}}+H_{f}
\end{equation}%
При $N=2$ мы возвращаемся к $s-f$ модели со спином $S=1/2$, а случай $M=N$ соответствует модели Коблина-Шриффера, где также имеет место полная компенсация. Более реалистическая модель, включающая угловые моменты, обсуждается в \cite{IK}; возможно и обобщение на случай произвольного спина $S$ (см, напр., \cite{Cox,507nfl}).

Обсудим различные возможности формирования магнитного состояния. Самым простым механизмом является возникновение режима недокомпенсации магнитных моментов ($S>M/2$ при $N=2$). По-видимому, он реализуется в некоторых ферромагнитных урановых соединениях, где эффект Кондо заметно проявляется только в парамагнитной фазе \cite{Perkins}.

Более обычна ситуации полной компенсации $S=M/2$, которая естественно  возникает в реальных системах со сложной электронной структурой и вырожденными зонами. Здесь относительные роли эффекта Кондо и межузельного РККИ-взаимодействия задаются
величинами двух энергетических масштабов: температуры Кондо
$T_{\text{K}}\sim D\exp (1/2I\rho )$ ($\rho= \rho(E_F)$ -- затравочная плотность состояний на уровне Ферми, $D$ – ширина зоны проводимости), которая определяет кроссовер
между режимом свободных моментов и областью сильной связи, и
$T_{\text{RKKY}} \sim I^2\rho $. Последняя величина имеет
порядок температуры магнитного упорядочения $T_{\text{M}}$ в
отсутствие эффекта Кондо. Отношение $T_{\text{K}}/T_{\text{M}}$
может изменяться в зависимости от внешних параметров и состава
системы при легировании.


В~немагнитном случае $T_{\text{RKKY}}\sim \overline{\omega }$, где
$\overline{\omega }$~"--- характерная частота cпиновых флуктуаций.
Для большинства обсуждаемых соединений
$T_{\text{K}}>T_{\text{RKKY}}$. Однако существуют также аномальные
магнетики, содержащие церий и уран, с $T_{\text{K}}\ll
T_{\text{N}}$, например CeAl${}_2$Ga${}_2$, UAgCu${}_4$.
Аналогичная ситуация имеет место в ферромагнитных системах CeRu$_2$X$_2$M \cite{CeRu2X2M1,CeRu2X2M2}.
Этот случай близок к элементарным редкоземельным магнитным металлам, где эффект Кондо практически полностью подавлен магнитным упорядочением и дает лишь малые поправки к магнитному моменту.

В качестве первого шага к описанию формирования состояния кондовского магнетика
полезно рассмотреть поправки теории возмущений к магнитным характеристикам~\cite{367,IK}. При этом, в отличие от однопримесной проблемы Кондо, следует учесть эффекты межпримесного обменного взаимодействия, которое приводит к возникновению спиновой динамики. Вычисление магнитной восприимчивости приводит к результату
\begin{equation}
\chi =\frac{S(S+1)}{3T}(1-4I^2L), \,
L=\frac{MN/2}{S(S+1)}\sum_{\mathbf{p}\mathbf{q}}\int
K_{\mathbf{p}-\mathbf{q}}(\omega )
\frac{n_{\mathbf{p}}(1-n_{\mathbf{q}})}
{(t_{\mathbf{q}}-t_{\mathbf{p}}-\omega )^2}\,d\omega
 \label{eq:6.83}
\end{equation}
где
$K_{\mathbf{q}}(\omega )$ -- спиновая спектральная плотность.

Из простой логарифмической оценки интеграла в~(\ref{eq:6.83}) следует
\begin{equation}
\chi =\frac{S(S+1)}{3T}\left( 1-MNI^2\rho ^2 \ln{}
\frac{D^2}{T^2+\overline{\omega }^2}\right) ,
 \label{eq:6.85}
\end{equation}
где величина в скобках описывает подавление эффективного момента.

Кондовские поправки к магнитному моменту в ферро- и
антиферромагнитном состояниях получаются с использованием
стандартного спин-волнового результата
\begin{equation}
\delta \overline{S}=-\sum_{\mathbf{q}}N_{\mathbf{q}}, \, N_{\mathbf{q}} = \langle b_{\mathbf{q}}^{\dagger
}b_{\mathbf{q}}^{}\rangle
 \label{eq:6.86}
\end{equation}
подстановкой поправки к числам заполнения магнонов $N_{\mathbf{q}}$ при нулевой температуре, которая обусловлена затуханием спиновых волн вследствие рассеяния на
электронах проводимости. В случае ферромагнетика имеем (здесь для простоты $N=2, M=1$)
\begin{equation}
\delta N_{\mathbf{q}}
=2I^2S\sum_{\mathbf{k}}\frac{n_{\mathbf{k}\downarrow }
(1-n_{\mathbf{k}-\mathbf{q}\uparrow })}{(t_{\mathbf{k}\downarrow }
-t_{\mathbf{k}-\mathbf{q}\uparrow }-\omega _{\mathbf{q}})^2},
 \label{eq:6.87}
\end{equation}
где $ n_{\mathbf{k}\sigma}$ -- числа заполнения электронов (фермиевские функции). Вычисление как для ферромагнетика, так и антиферромагнетика дает \cite{IK,I16}
\begin{equation}
\delta \overline{S}/S=-MNI^2\rho ^2\ln{} \frac{D}{\overline{\omega }}.
 \label{eq:6.89}
\end{equation}
Эти поправки к моменту в основном состоянии возникают в любых
проводящих магнетиках, включая чистые $f$"~металлы. Однако в
последнем случае они должны быть малы (порядка $10^{-2}$).

Поправки к электронным характеристикам могут быть вычислены анлогичным образом; в электронной собственной энергии $\Sigma _{\mathbf{k}} (E)$ кондовские вклады третьего порядка (определяющие перенормировку $s-f$ обменного параметра) также обрезаются на $\overline{\omega }$.
Следует отметить, что при наличии спиновой динамики расходимости кондовского типа в $\Sigma _{\mathbf{k}} (E)$ возникают уже в членах второго порядка, которые дают вклад в усиление эффективной массы \cite{367}. Формально эти расходимости связаны с функцией Ферми, а соответствующие вклады аналогичны электрон-фононной или спин флуктуационной (парамагнонной) перенормировке.

В~целях получения самосогласованной картины для магнетика с
заметными кондовскими перенормировками нужно вычислить
поправки к характерным частотам спиновых флуктуаций
$\overline{\omega }$. В~парамагнитной фазе
оценка  поправки второго порядка к динамической восприимчивости
дает~\cite{608}:
\begin{equation}
\omega _{\mathbf{q}}^2=\frac{4}{3}S(S+1)\sum_{\mathbf{p}}
(J_{\mathbf{q}-\mathbf{p}}-J_{\mathbf{p}})^2 [1-4I^2L(1-\alpha
_{\mathbf{q}})].
 \label{eq:6.92}
\end{equation}
Здесь величина $L$ определена в~(\ref{eq:6.83}),
\begin{equation}
\alpha _{\mathbf{q}}=\sum_{\mathbf{R}}J_{\mathbf{R}}^2\left(
\frac{\sin{} k_{\text{F}}R}{k_{\text{F}}R}\right) ^2[1-\cos{}
\mathbf{q}\mathbf{R}]\Biggm/ \sum_{\mathbf{R}}
J_{\mathbf{R}}^2[1-\cos{} \mathbf{q}\mathbf{R}],
 \label{eq:6.93}
\end{equation}
сумма идет по векторам решетки. Поскольку $0<\alpha _{\mathbf{q}}<1$, эффект Кондо приводит к
уменьшению $\overline{\omega }(T)$ при понижении
температуры. В~приближении ближайших соседей (с периодом решетки
$d$) для $J(\mathbf{R})$ значение~$\alpha $ не~зависит от~$\mathbf{q}$, так что
$\alpha = \sin^2(k_{\text{F}}d)/(k_{\text{F}}d)^2$.
Вычисление поправок к частоте спиновых волн в ферромагнитной и
антиферромагнитной фазе вследствие магнон-магнонного
взаимодействия, а также электрон-магнонного рассеяния приводит к результату
\cite{IK,I16}
\begin{equation}
\delta \omega _{\mathbf{q}}/\omega _{\mathbf{q}}=-2MNI^2\rho
^2a\ln{} \frac {D}{\overline{\omega }},
 \label{eq:6.95}
\end{equation}
где множитель $a$ зависит от типа магнитного упорядочения.

Приведенные результаты теории возмущений дают возможность
качественного описания состояния кондо-решетки как магнетика с
малым магнитным моментом \cite{608}. Будем понижать температуру, стартуя с парамагнитного состояния. При этом
магнитный момент "<компенсируется">, но, в отличие от
однопримесной ситуации, степень компенсации определяется
$(T^2+\overline{\omega }^2)^{1/2}$ вместо~$T$. В~то~же время сама
$\overline{\omega }$ уменьшается согласно~(\ref{eq:6.92}). Этот
процесс не~может быть описан аналитически в рамках теории
возмущений. Однако, если иметь в виду образование единого универсального
энергетического масштаба порядка $T_{\text{K}}$, то нужно выбрать
$\overline{\omega }\sim T_{\text{K}}$ при $T<T_{\text{K}}$.
Последний факт подтверждается большим числом экспериментальных
данных относительно квазиупругого нейтронного рассеяния в
кондо-системах, которые показывают, что при низких температурах
типичная ширина центрального пика $\Gamma \sim \overline{\omega }$
имеет тот~же самый порядок величины, что и фермиевская температура
вырождения, определенная из термодинамических и кинетических
свойств, т.~е. $T_{\text{K}}$. Следовательно, процесс компенсации
магнитного момента завершается где-то на границе области сильной
связи и приводит к состоянию с конечным (хотя, возможно, и малым)
моментом насыщения~$\mu_s$.

Более последовательное рассмотрение проблемы магнетизма решеток Кондо
может быть выполнено в рамках подхода ренормгруппы. В простейшей форме андерсоновского "<скейлинга для бедных"> (poor man scaling)~\cite{554} ренормгрупповые уравнения для
эффективного $s-f$~параметра и $\overline{\omega
}$~записываются на основе рассмотрения интегралов по
импульсам в многоэлектронных кондовских поправках к электронной собственной энергии и частоте спиновых флуктуаций \cite{612,IK,IK11}.

Чтобы построить процедуру скейлинга,
нужно выделить вклады от энергетического слоя $C<E<C+\delta C$
около уровня Ферми $E_{\text{F}}=0$. Например, в случае
ферромагнетика из эффективного расщепления в электронном спектре, определяемого из собственной энергии второго порядка,
\begin{equation}
2I_{\text{eff}}S=2IS-\left[ \Sigma _{\mathbf{k}\uparrow
}^{\text{FM}}(E_{\text{F}})-\Sigma _{\mathbf{k}\downarrow
}^{\text{FM}}(E_{\text{F}})\right] _{k=k_{\text{F}}}
 \label{eq:6.96}
\end{equation}
находим
\begin{equation}
\delta
I_{\text{ef}}=I^2\sum_{C<t_{\mathbf{k}+\mathbf{q}}<C+\delta
C}\left( \frac {1}{t_{\mathbf{k}+\mathbf{q}}+\omega _{\mathbf{q}}}
+\frac{1}{t_{\mathbf{k}+\mathbf{q}}-\omega _{\mathbf{q}}}\right)=
\frac{\rho I^2}{\overline{\omega }}\delta C\ln{} \left| \frac
{C-\overline{\omega }}{C+\overline{\omega }}\right|,
 \label{eq:6.97}
\end{equation}
где $\overline{\omega }=4 \mathcal{D}k_{\text{F}}^2$,
$\mathcal{D}$~"--- спин-волновая жесткость.
Вводя безразмерные константы
связи $g=-NI\rho, g_{\text{ef}}(C)=-NI_{\text{ef}}(C)\rho$ и заменяя в поправках теории возмущений $g \rightarrow g_{\text{ef}}(C)$,
получаем систему уравнений ренормгруппы в однопетлевом приближении:

\begin{equation}
\partial g_{\text{ef}}(C)/\partial C=-\Lambda , \qquad
\partial\ln{} \overline{\omega }_{\text{ef}}(C)/\partial C=aM\Lambda /N, \qquad
\partial\ln{} \bar{S}_{\text{ef}}(C)/\partial C=M\Lambda /N,
 \label{eq:6.99}
\end{equation}
где
\begin{equation}
\Lambda =\Lambda (C,\overline{\omega }_{\text{ef}}(C))
=[g_{\text{ef}}^2(C)/C]\eta (\overline{\omega
}_{\text{ef}}(C)/C).
 \label{eq:6.100}
\end{equation}
Скейлинговая функция для пара-, ферро- и антиферромагнитных фаз
равна:
\begin{equation}
\eta (x)=
\begin{cases}
x^{-1}\arctan{} x,                               & \text{PM}, \\ %
\frac{1}{2x}\ln{} \left| \frac{1+x}{1-x}\right|, & \text{FM}, \\ %
-x^{-2}\ln|1-x^2|,                               & \text{AFM}.   %
\end{cases}
 \label{eq:6.101}
\end{equation}
Как показывают результаты исследования
уравнений~(\ref{eq:6.99})---(\ref{eq:6.101})~\cite{612,IK},
в~зависимости от соотношения между однопримесной температурой
Кондо и затравочной частотой спиновых флуктуаций возможны три
режима:
\begin{enumerate}
    \item Режим сильной связи, где $g_{\text{ef}}$ расходится
при некотором $C$. Он грубо определен условием
$\overline{\omega }<T_{\text{K}}=D\exp{} (-1/g)$.
Здесь $I_{\text{ef}}(C\rightarrow 0)=-\infty $, так что все
электроны проводимости связаны в синглетные состояния и спиновая
динамика подавлена.
    \item Режим "<кондовского"> магнетика с заметной, но
неполной компенсацией магнитных моментов, который реализуется в
интервале $T_{\text{K}}<\overline{\omega }<AT_{\text{K}}$
($A$~"--- числовой множитель порядка единицы), соответствующем
малому интервалу $\delta g\sim g^2$. В~этом интервале
перенормированные значения магнитного момента и частоты спиновых
флуктуаций $\bar{S}_{\text{ef}}(0)$ и $\overline{\omega
}_{\text{ef}}(0)$ увеличиваются от нуля  почти до затравочных
значений.
    \item Режим "<обычных"> магнетиков с малыми
логарифмическими поправками к моменту основного состояния
(см.~(\ref{eq:6.89})), возникающий при $\overline{\omega
}>AT_{\text{K}}$. В этом случае нет существенной зависимости от знака $I$.
\end{enumerate}
В случае~2 имеет место высокая чувствительность магнитного состояния к внешним факторам, так что
магнитный момент сильно меняется при малых вариациях затравочного
параметра взаимодействия.

Переход в магнитное состояние по затравочной константе связи $g$ описывается как квантовый фазовый переход со своими критическими индексами \cite{IK}. В его узкой окрестности возникает НФЖ поведение, где $g_{\text{ef}}$ возрастает линейно с увеличением $\xi = \ln|D/C|$ (см. обсуждение в следующем разделе).
Критическое значение $g_c$ существенно зависит от типа магнитного
упорядочения и структуры магнонного спектра (например, наличия в
нем щели), размерности пространства и др. Таким образом, критерий
Дониаха $g_c\simeq 0.4$ \cite{Don},
полученный для простой одномерной модели,
вряд ли может быть применимым к
реальным системам. В рамках первопринципных расчетов эти проблемы
обсуждаются в работе \cite{Don1}.

Разумеется, при рассмотрении квантового магнитного фазового
перехода требуется более точный учет магнитных флуктуаций, включая импульсные зависимости. Недавно
для анализа фазовой диаграммы двумерной антиферромагнитной
решетки Кондо было использовано $\epsilon$-разложение в методе
ренормгруппы с использованием нелинейной сигма-модели \cite{2Drg}.
Важную роль также может играть изменение топологии поверхности Ферми
\cite{Si}.

\section{Магнетизм и нефермижидкостное поведение. Многоканальная модель Кондо}


Для объяснения нефермижидкостного поведения  $f$-систем был
выдвинут ряд идей.
Поскольку многие НФЖ соединения являются неупорядоченными сплавами, были предложены механизмы, связанные с разупорядочением в решетках Кондо \cite{Miranda}, а также использована модель сингулярностей Гриффитса--Mак-Коя \cite{Griffiths}.
Как упомянуто во Введении, НФЖ поведение чаще всего наблюдается на границе магнитной неустойчивости -- при квантовых фазовых переходах (QPT), в непосредственной близости которых (регулируемой составом либо внешним давлением) магнитный порядок подавлен вплоть до  нулевой температуры.
В этой связи предлагались механизмы, связанные со спиновыми флуктуациями \cite{26}, и подходы, рассматривающие поведение вблизи  QPT,  --  в "<чистом"> пределе  \cite{Millis} или с учетом разупорядочения \cite{Belitz}.

Как показывают расчеты по теории возмущений в $s-f$ модели антиферромагнетика \cite{afm}, в двумерном случае ($d=2$) или в ситуации фрустрированного магнонного спектра для $d=3$ межзонные вклады в электронную теплоемкость дают зависимость $T \ln(T/T^*)$ вместо линейной, а вместо квадратичной зависимости сопротивления имеем при $d=2$ либо при $d=3$ в ситуации "<нестинга"> для электронного спектра имеем $R(T) \sim T \ln(T/T^*)$. Разумеется, эти зависимости справедливы в ограниченном температурном интервале $T>T^*$, где величина $T^* \sim T_N \bar{\Delta} /E_F$ определяется величиной спинового расщепления $\bar{\Delta} = 2|I|S $ (в случае ферромагнетика $ T^*$ порядка $T_C I^2 $ из-за квадратичного спектра спиновых волн). Аналогичные зависимости возникают в слабых коллективизированных моделях на границе магнитного фазового перехода \cite{Kampf,Moriya,26}.
С другой стороны, как мы видели, в некоторых веществах НФЖ поведение сохраняется существенно выше температуры упорядочения или  в магнитоупорядоченной фазе – достаточно далеко от критической точки QPT.

НФЖ поведение может быть описано в многоканальной модели Кондо \cite{Tsv,Col,Gan,Cox}, где оно возникает в режиме переэкранирования магнитных моментов. Здесь состояние полной компенсации неустойчиво: с локализованным спином (даже при $S=1/2$) связывается целый комплекс электронов проводимости, и получающийся композитный объект продолжает взаимодействовать с морем электронов. В результате возникает инфракрасная фиксированная точка.

В данной  ситуации необходимо обобщить ренормгрупповое рассмотрение предыдущего раздела, рассматривая скейлинг высшего порядка. В однопримесном случае скейлинговое поведение определяется бета-функцией
$\beta (g)= \partial g_{\text{ef}}(C)/\partial \ln |C| $,
разложение которой дает \cite{Gan}
\begin{equation}
\beta (g)= -g^{2}+(M/N)g^{3}+...
\end{equation}%
При $M>N$ (для простоты обсуждается случай $S=1/2$) фиксированная точка $g^{\ast }=N/M$ (нуль $\beta (g)$) лежит в области слабой связи, так что оправдано использование теории возмущений и ренормгруппы.
Напротив, в случае компенсации ($M=1$), обсуждавшемся в разделе 3, такое рассмотрение неприменимо – фиксированная точка с большим $g^{\ast } $ является нефизической.

Решение скейлингового уравнения дает
\begin{equation}
\frac{g^{\ast }-g_{ef}(C)}{g^{\ast }-g}=g^{\ast }\left\vert \frac{C}{T_{K}}%
\right\vert ^{\Delta }\exp \left( -\frac{g^{\ast }}{g_{ef}(C)}\right)
\label{pow}
\end{equation}%
где $\Delta =N/M$ и температура Кондо равна
\begin{equation}
T_{K}=Dg^{M/N}\exp (-1/g)  \label{TK}.
\end{equation}%
В таком подходе  расходимость $g_{ef}(\xi )$ отсутствует, а степенное критическое поведение (\ref{pow}) имеет место в широком интервале, включая область $|C|>T_{K}$ \cite{Gan}.
Критические показатели определяются наклоном $\Delta =\beta^{\prime }(g).$ Вычисление с учетом высших порядков по $1/M$ приводит к результату
\begin{equation}
\Delta =\frac{N}{M}\left( 1-\frac{N}{M}\right) \simeq \frac{N}{M+N},
\label{Delts}
\end{equation}%
что согласуется с точными результатами из анзатца Бете и конформной теории поля
 (см. \cite{Cox}). Для соответствующего значения $g^{\ast }$ при $N=2$ разложение дает $g^{\ast }=\frac{2}{M}\left( 1-\frac{2\ln 2}{M}\right)$ \cite{Gan},
что слегка отличается от $\Delta .$

Важный случай $M = N=2 $ более сложен с теоретической точки зрения и не может быть рассмотрен простыми аналитическими методами. Здесь восприимчивость и коэффициент теплоемкости при понижении температуры расходятся логарифмически, как дают те же анзатц Бете и конформная теория поля (см. обзор  \cite{Cox}).

Перейдем теперь к случаю решетки  \cite{I16}. Аналогично рассмотрению в разделе 3 находим
\begin{equation}
\partial g_{ef}(C)/\partial C =-[1-(N/M) g_{ef}(C)] \Lambda
\label{sef}
\end{equation}%
При этом перенормировка частоты $\overline{\omega }$ в ведущем порядке по $M$ описывается тем же уравнением, что и в (\ref{eq:6.99}). Тогда
имеем ($\gamma =M/N$)
\begin{equation}
\frac{\overline{S}_{ef}(C)}{S} =\left( \frac{\overline{\omega }_{ef}(C)}{%
\overline{\omega }}\right) ^{1/a}=\frac{1-\gamma g_{ef}(C)}{1-\gamma g}=%
\frac{g^{\ast }-g_{ef}(C)}{g^{\ast }-g},  \label{w+g}
\end{equation}%
где $a$ введено в (10). Таким образом, мы имеем ситуацию мягкой бозонной моды при приближении к фиксированной точке $g^*=1/\gamma $.

Вводя функцию
\begin{equation}
\psi (\xi )=\ln \frac{\overline{\omega }}{\overline{\omega }_{ef}(\xi )}%
=a\ln \frac{\overline{S}_{ef}(C)}{S}
\end{equation}%
 можно записать скейлинговое уравнение в виде
\begin{equation}
\frac{\partial \psi }{\partial \xi }=\frac{a}{\gamma }\left[ 1-(1-\gamma
g)\exp (-\psi  /a)\right] ^{2}\Psi (\lambda +\psi  -\xi )  \label{linf}
\end{equation}%
где
\begin{equation*}
\Psi (\xi )=\eta (e^{-\xi }),\qquad \xi =\ln |D/C|,\qquad \lambda =\ln (D/\overline{%
\omega })\gg 1
\end{equation*}%
Приведенные уравнения записаны в терминах $ \gamma $, а не $ M $ и $ N $ по отдельности. Поэтому, чтобы установить соответствие с однопримесным случаем (\ref{Delts}), можно положить  $\gamma =M/N+1=1/\Delta .$ Это дает, по крайней мере при $ M> 2 $, правильные критические показатели для магнитной восприимчивости, теплоемкости и сопротивления.

\begin{figure}[tbp]
\includegraphics[width=3.3in, angle=0,clip]{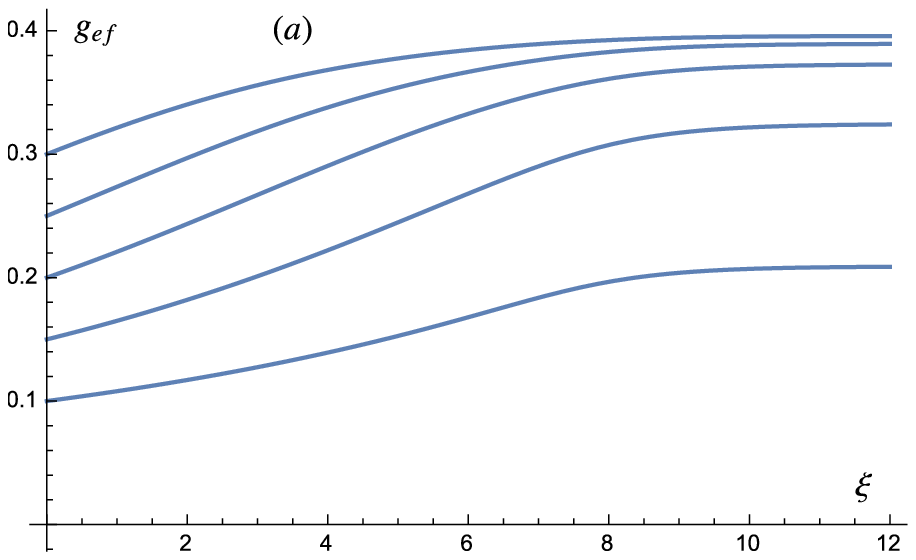}
\includegraphics[width=3.3in, angle=0,clip]{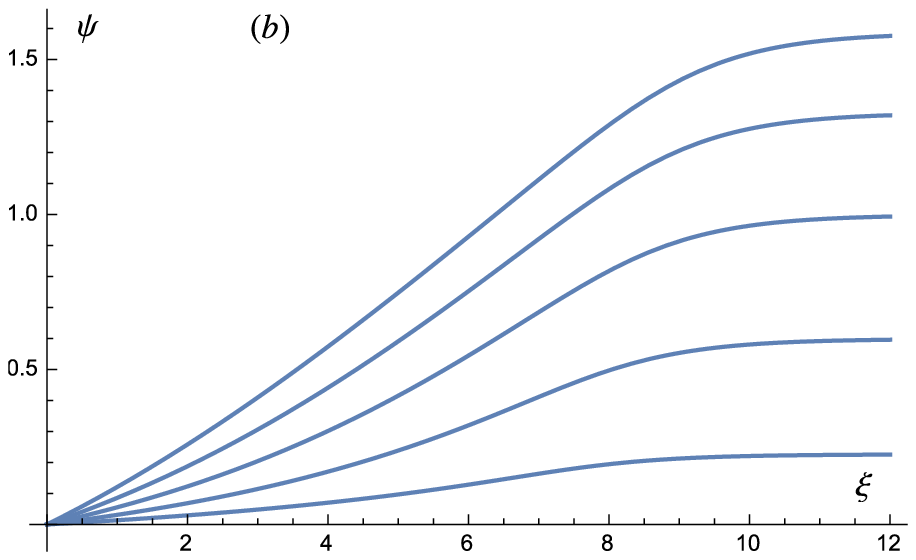}
\caption{Скейлинговые траектории для случая парамагнетика $g_{ef}(\xi )$ (a)
и $\psi (\xi )$ (b) \cite{I16}. Значения параметров: $\gamma = 5/2$, $\lambda =7$, $a=1/2$, $g=0.1, 0.15, 0.2, 0.25, 0.3$ (снизу вверх)}
\label{fig:1}
\end{figure}

В  парамагнитной фазе при больших $\xi =\ln|D/C|$ можно положить для оценки $g_{ef}^{{}}(\xi )\simeq g^{\ast }=1/\gamma $.  Тогда имеем степенное поведение
\begin{eqnarray}
\overline{\omega }_{ef}(C) &\simeq &\overline{\omega }(|C|/T_{K})^{\beta
},\quad \beta =a/\gamma =a\Delta  \nonumber \\
\overline{S}_{ef}(C) &\simeq &(|C|/T_{K})^{\Delta },~  \label{pm}
\end{eqnarray}%
Оно имеет место в ограниченном интервале значений $\xi$, определяемом спиновой динамикой, --- вплоть до $\xi =\xi _{1}\simeq (\lambda -\beta /g)/(1-\beta )$. При $\xi \rightarrow \infty $ величина $\psi (\xi )$ остается конечной, а $g_{ef}(\xi )$ стремится к значению, несколько меньшему, чем однопримесное $g^{\ast }$ (рис. \ref{fig:1}).

\begin{figure}[tbp]
\includegraphics[width=3.3in, angle=0]{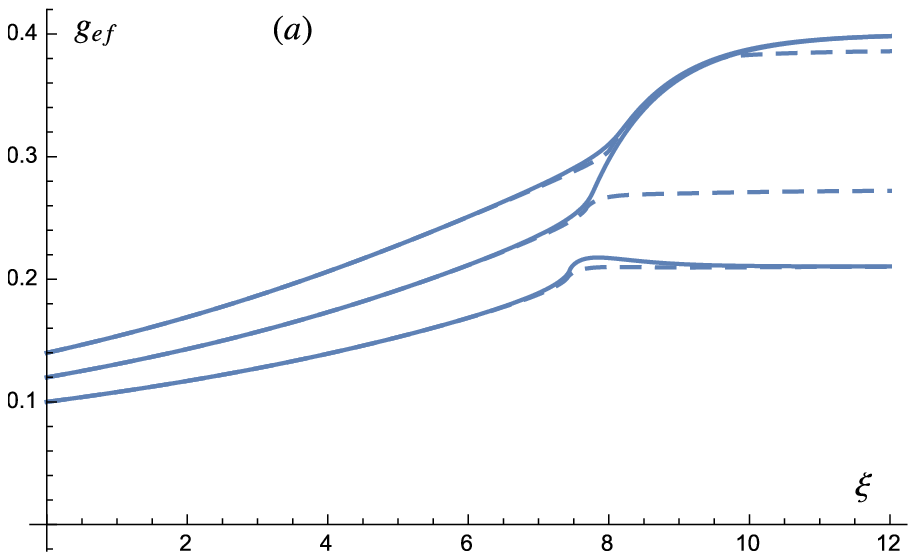}
\includegraphics[width=3.3in, angle=0]{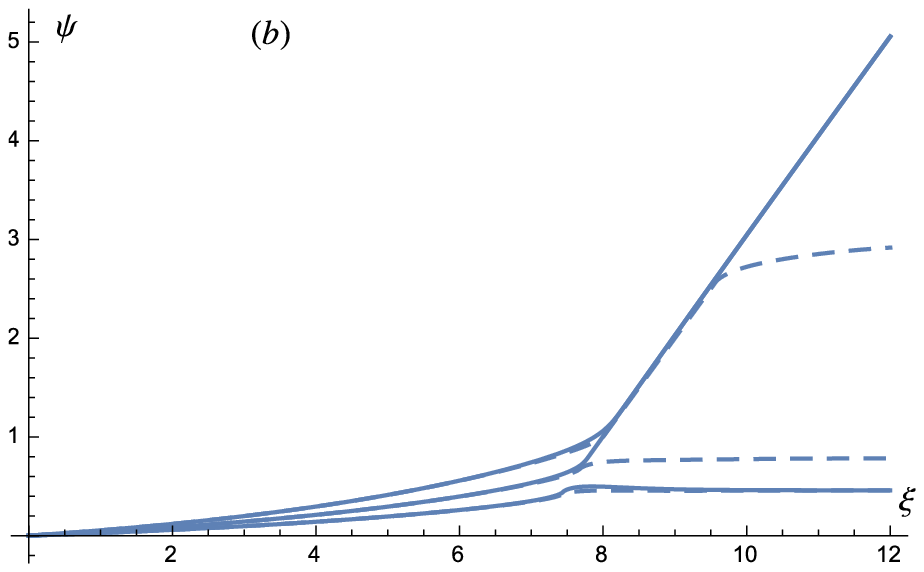}
\caption{
Скейлинговые траектории для случая трехмерного антиферромагнетика $g_{ef}(\xi )$ (a)
и $\psi (\xi )$ (b) \cite{I16} для $\gamma = 5/2$, $\lambda =7$, параметра затухания $\delta =2~10^{-4},$ $a=1 $, $g=0.1,0.12,0.14$ (снизу вверх). Пунктир соответствует учету некогерентного вклада в спиновую спектральную плотность
}
\label{fig:2}
\end{figure}

В случае магнитного упорядочения поведение для $\xi <\xi _{1}$ аналогично, но для $\xi >\xi _{1}$ важную роль играют сингулярности Ван-Хова в скейлинговой функции: вместо убывания $\Psi (\lambda +\psi -\xi )$ может начать возрастать при приближении к $\xi _{1}$,
аргумент функции $\Psi$ in (\ref{linf}) становится практически постоянным (фиксированным), и мы имеем (Рис. \ref{fig:2})
\begin{equation}
\psi (\xi )\simeq \xi -\lambda ,\;\overline{\omega }_{ef}(C)\simeq |C|.
\label{lin}
\end{equation}%
В отличие от парамагнитного случая, здесь имеет место резкий кроссовер с изменением  $g$: режим (\ref{lin}) не достигается при $g<g_{c}$. Значение $g_{c}$ определяется величиной магнонного затухания $\delta ,$, на котором обрезаются сингулярности функции (\ref{eq:6.101}).
Поведение (\ref{lin}) имеет критическую природу: в случае одноканальной модели Кондо оно возникает только при критическом $g=g_{c}$, соответствующем квантовому магнитному фазовому переходу.

Вообще, однопетлевой скейлинг, рассмотренный в предыдущем разделе, дает для конечных $M$ НФЖ поведение лишь в очень узком интервале затравочной константы связи $g$, поскольку с ростом $g$ мы быстро переходим в режим сильной связи, где $g_{ef}(\xi >\lambda)\rightarrow \infty .$
Однако существуют механизмы, приводящие к расширению области НФЖ поведения и в случае $M=1$ (когда описанный механизм в данном разделе возникновения фиксированной точки не реализуется). В частности, это происходит при учете затухания спиновых возбуждений ~\cite{731}.

Как показано в~\cite{731}, в случае особого вида спектральной функции спиновых возбуждений, который может реализоваться в многоподрешеточных магнетиках, аргумент функции $\Psi $ также может фиксироваться на скейлинговой траектории. Это приводит к линейному возрастанию $g_{ef}(\xi)$, так что НФЖ поведение возникает в широком интервале затравочной константы связи $ g_{c1}<g<g_{c2}$,

При наличии ван-хововских особенности электронного спектра область НФЖ поведения также расширяется \cite{I11}.
Такие особенности могут существенно повлиять на структуру теории возмущений. При этом скейлинговое поведение будет определяться не только близостью уровня Ферми ($C \rightarrow 0$), но и расстоянием  до особенности;
 в определенных ситуациях возможно ослабление зависимости константы связи от ее затравочного значения.
Интересной возможностью является пиннинг (залипание) уровня Ферми у особенности
ван-Хова \cite{pinning}.


Существует также возможность, что фиксированная точка, аналогичная случаю многоканальной модели Кондо, появится в случае одноканальной модели с особенностью плотности состояний или нестинга \cite{I17} (отметим, что гигантские особенности ван-Хова формируются при пересечении более слабых особенностей \cite{81}, т.е. связаны с вырождением зон).

Теперь обсудим поведение физических свойств для случая $ N = 2.$ Температурные зависимости магнитного момента и восприимчивости в парамагнитном случае получаются непосредственно из приведенных выше результатов с помощью замены $ |C| \rightarrow T $,%
\begin{equation}
S_{ef}^{{}}(T)\varpropto (T/T_{K})^{\Delta },~\chi(T)\varpropto
S_{ef}^{2}(T)/T\varpropto (T/T_{K})^{2\Delta }/T.
\end{equation}%
Аналогичная зависимость получается для электронной теплоемкости \cite{Gan}.
В отличие от однопримесного случая, эти зависимости справедливы в ограниченном темературном интревале.

Температурная поправка к магнитному сопротивлению имеет вид
\cite{Gan}%
\begin{equation}
\delta R_{m}(T)\propto g_{ef}(T)-g^{\ast }\propto -(T/T_{K})^{\Delta }.
\end{equation}%
Зависимость $T^{1/2}$ (которая соответствует $M=2$) действительно наблюдается в ряде $f$-систем \cite{Cox}. С другой стороны, как обсуждалось выше, рассматриваемое приближение не учитывает логарифмических расходимостей, возникающих в восприимчивости и теплоемкости при $M=2$ ($\Delta =1/2$).

В спин-волновой области имеем для антиферромагнетика
$\chi \propto \overline{S}/\overline{\omega }.$
Заменяя $\overline{\omega }\rightarrow \overline{\omega }_{ef}(C),$ $%
\overline{S}\rightarrow \overline{S}_{ef}(C)$ с $|C|\sim T$ в духе скейлингового рассмотрения, находим
\begin{equation}
\chi (T)\propto T^{-\zeta },~\zeta =\left\{
\begin{array}{c}
\Delta (a-1)/a \\
(a-1)/a%
\end{array}%
\right.
\end{equation}%
в режимах (\ref{pm}) и (\ref{lin}) соответственно.
Неуниверсальный показатель $\zeta $ определяется деталями магнитной структуры и может быть как положительным, так и отрицательным.

Для электронной теплоемкости в антиферромагнитном случае имеем
\begin{equation}
C_{el}(T)/T\propto g_{ef}^{2}(T)\overline{S}_{ef}(T)/\overline{\omega }%
_{ef}(T)\propto \chi (T).
\end{equation}%
что находится в согласии с экспериментальными данными по НФЖ системам \cite{Tsv,Cox}.

Таким образом, многоканальная модель решетки Кондо естественно описывает формирование магнитного состояния с малым значением момента. Кроме того, она дает пример существенной перенормировкой константы связи в соответствии с (\ref{w+g}). Это может представлять интерес для общей теории металлического магнетизма (в частности, для слабого зонного ферро- и антиферромагнетиках): магнитное состояние определяется процессом перенормировки, а не критерием Стонера с затравочными параметрами.

Важной проблемой является устойчивость фиксированной точки: снятие вырождения электронных подзон с различным $ m$ в гамильтониане (\ref{eq:G.2}) должно приводить к изменению скейлингового поведения, так что аномальные температурные зависимости будут сохраняться лишь в ограниченном интервале. Применения двухканальной модели к редкоземельным и актинидным системам, в том числе соответствующие трудности интерпретации, обсуждаются в обзоре \cite{Cox}. В случае урановых систем реализация этой модели возможна благодаря симметрии подзон относительно инверсии времени.

\section{Проблема основного состояния. Приближение среднего поля в представлении псевдофермионов}

Методы теории возмущений и скейлинга, рассмотренные выше, не позволяют описать основное состояние решеток Кондо в режиме сильной связи (в этой связи следует отметить работу \cite{Barnes}, где такое описание было получено в однопримесной модели диаграммным методом). Поэтому проблема сильно коррелированного основного состояния плотных кондовских систем весьма сложна.

Принципиальным является тот факт, что $f$"~-электроны в системах с тяжелыми фермионами ведут себя необычно – проявляют делокализацию,
которая подтверждена наблюдением больших электронных масс в
экспериментах по эффекту де~Гааза"--~ван~Альфена.
Такое поведение весьма нетривиально в
$s-f$~обменной модели (в отличие от модели Андерсона с
$f$"~состояниями вблизи уровня Ферми, где имеется затравочная
$s$---$f$~гибридизация и возникает состояние промежуточной валентности).
Эта делокализация аналогична появлению
фермиевской ветви возбуждения в теории резонирующих валентных
связей для высокотемпературных сверхпроводников, так что поверхность Ферми может быть  спинонной, причем не только в двумерном, но и в трехмерном случае \cite{Sachdev}.
В результате коллективизации $f$-состояний нарушается теорема Латтинджера о постоянстве объема под поверхностью Ферми и возникает так называемая большая поверхность Ферми (это изменение статистики в чем-то аналогично формированию и разрушению хаббардовских подзон \cite{Hubbard-I:1963}).

В основном состоянии при изменении параметра взаимодействия происходит квантовый фазовый переход в металл с обычной малой поверхностью Ферми.
Такой скачок поверхности Ферми действительно наблюдается в YbRh$_2$Si$_2$ и CeRhIn$_5$~\cite{Si}.
При этом остается вопрос, как формируется магнетизм: возникает ли магнитное упорядочение при разрушении кондовского состояния, или магнитный переход происходит в границах кондовской фазы путем формирования волны спиновой плотности (SDW). В последнем случае формирование малых моментов происходит естественно, по крайней мере в ситуации антиферромагнетизма. В то же время магнитное упорядочение, меняя зону Бриллюэна, само может оказывать существенное влияние на зонную структуру и поверхность Ферми.

\begin{figure}[tbp]
\includegraphics[width=3.3in, angle=0,clip]{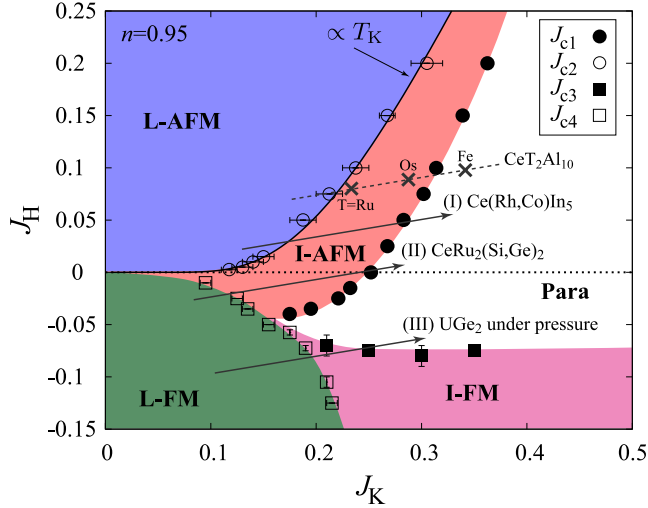}
\caption{Фазовая диаграмма решетки Кондо--Гейзенберга в плоскости $J_{\rm H}$--$J_{\rm K}$(где $J_{\rm K} =-2I$) при $T=0.001$ \cite{Hoshino}. Сокращения L и I означают локализованный и коллективизированный (itinerant);
величины по осям в единицах ширины зоны}
\label{fig_gs}
\end{figure}

Переход от малой поверхности Ферми к большой может рассматриваться как переход от локализованного к коллективизированному магнетизму.
Соответствующее обсуждение на основе расчетов методом Монте-Карло в комбинации с методом динамической теории среднего
поля (DMFT) проведено в работе \cite{Hoshino} в применении к фазовым переходам в системах CeRh$_{1-x}$Co$_{x}$In$_{5}$, CeRu$_{2}$ (Si$_{x}$Ge$_{1-x}$)$_{2}$, UGe$_{2}$, CeT$_{2}$Al$_{10}$ (T = Fe, Ru, Os), причем было рассмотрено как ферро-, так и антиферромагнитное гейзенберговское взаимодействие (рис.~\ref{fig_gs}).

Конкуренция магнитных взаимодействий (фрустрации) может приводить к многообразию фазовых переходов, в том числе между магнитными и парамагнитными состояниями  с большой и малой поверхностью Ферми (см. обсуждение в \cite{Si,frustr,Isaev}). Например, в Ce$_3$Pd$_{20}$Si$_6$ и YbRh$_2$Si$_2$ кондовское экранирование, по-видимому, исчезает при магнитном переходе, а при легировании последнего соединения кобальтом или иридием возникают два отдельных фазовых перехода \cite{Friedemann}.
Согласно нейтронным данным для CeCu$_2$Si$_2$, магнитное упорядочении связано с нестингом большой поверхности Ферми тяжелых квазичастиц \cite{CeCu2Si2}.
Экзотическое магнитное состояние в CePdAl, обусловленное фрустрациями, обсуждается в \cite{CePdAl}.
Роль фрустраций в физике кондовских решеток ранее обсуждалась в работах \cite{PL,607}.

\begin{figure}[tbp]
\includegraphics[width=3.3in, angle=0,clip]{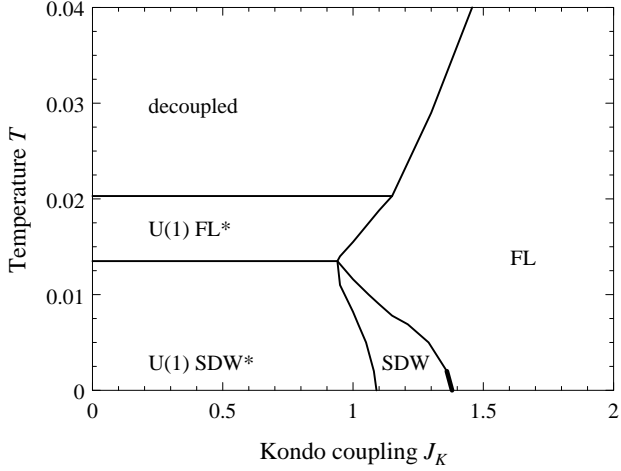}
\caption{
Магнитная фазовая диаграмма в приближении среднего поля из работы \cite{Sachdev}; decoupled --- фаза, в которой s- и f-электроны не связаны, $J_K = -2I$, величины по осям в единицах интеграла переноса
} \label{pht1}
\end{figure}

В работе \cite{Sachdev} был предложен новый тип ферми-жидкостного состояния FL$^*$. В  нем локализованные моменты не принимают участие в формировании поверхности Ферми, но адиабатически связаны со спиновой жидкостью, описываемой калибровочной теорией и обладающей соответствующими экзотическими возбуждениями в фазе деконфайнмента. В пространственных размерностях $d \geq 2$ стабильна спиновая жидкость типа  $Z_2$, однако более интересна спиновая жидкость U(1), существующая при  $d \geq 3$. В этой фазе коэффициент электронной теплоемкости $ C/T$ расходится логарифмически.
На фоне такого состояния возникает металлическое магнитное состояние SDW$^*$, которое может характеризоваться малым моментом. При этом магнитная неустойчивость развивается как волна спиновой плотности для спинонной поверхности Ферми.
С увеличением константы связи происходит переход в обычную (но тяжелую) ферми-жидкость FL (фазовая диаграмма показана на рис. \ref{pht1}).

В \cite{Sachdev} было искусственно введено поле, стабилизирующее ненасыщенное состояние с малым моментом.  В работе \cite{Isaev} за счет некоторой модификации модели и физической картины такого введения удалось избежать и было получено состояние, в котором эффект Кондо сосуществует с антиферромагнетизмом, хотя и при нереалистически большом параметре $s-f$ обмена.

Перейдем теперь к математическому формализму. В основном состоянии аномальных $f$-систем кондовскую компенсацию (образование синглета) следует рассматривать в нулевом приближении.
Для описания основного состояния кондо-решетки в режиме сильной связи было разработано специальное приближение среднего поля, в рамках которого Коулмен и Андрей~\cite{711}
рассмотрели формирование состояния спиновой жидкости в двумерной
ситуации. Затем это приближение было применено к случаю ферро- и антиферромагнитного упорядочения \cite{IK90,608,I14,Sachdev}.
Оно использует псевдофермионное представление для операторов  локализованных спинов $S=1/2$:
\begin{equation}
\mathbf{S}_i=\frac 12\sum_{\sigma \sigma ^{\prime }}f_{i\sigma
}^{\dagger }\mbox{\boldmath$\sigma $}_{\sigma \sigma ^{\prime
}}f_{i\sigma ^{\prime }}
 \label{eq:O.1}
\end{equation}
с вспомогательным условием
\begin{equation}
f_{i\uparrow }^{\dagger }f_{i\uparrow }+f_{i\downarrow }^{\dagger
}f_{i\downarrow }=1.
\label{eq:O.11}
\end{equation}
Ранее близкая идея о введении псевдофермионов была предложена в работе  \cite{Lacroix0}.
Варианты, связанные с использованием вспомогательных бозонов \cite{Moller}, оказываются существенно менее удачными, поскольку оправданы только в пределе большой кратности вырождения $N$  (при конечных $N$ и особенно при $N=2$ они дают экспоненту в выражении для температуры Кондо, отличную от однопримесного значения). Аналогичная проблема возникает в методе типа Гутцвиллера \cite{Fazekas}.

Применяя приближение перевальной точки для интеграла по траекториям, описывающего спин-фермионную взаимодействующую систему~\cite{711}, можно свести гамильтониан $s$---$f$~обменного взаимодействия к эффективной гибридизационной модели:
\begin{equation}
-I\sum_{\sigma \sigma ^{\prime }}c_{i\sigma }^{\dagger }c_{i\sigma ^{\prime }
}\left( \mbox{\boldmath$\sigma $}_{\sigma \sigma ^{\prime
}}\mathbf{S}_i-\frac 12\delta _{\sigma \sigma ^{\prime }}\right)
\rightarrow f_i^{\dagger }V_ic_i+c_i^{\dagger }V_i^{\dagger
}f_i-\frac 1{2I}\operatorname{Sp} (V_iV_i^{\dagger }),
 \label{eq:O.2}
\end{equation}
где введены векторные обозначения
$f_i^{\dagger }=(f_{i\uparrow }^{\dagger },\,f_{i\downarrow
}^{\dagger }),\, c_i^{\dagger }=(c_{i\uparrow }^{\dagger},
c_{i\downarrow }^{\dagger })$, а
$V$~"--- эффективная матрица гибридизации, определяемая из условия
минимума свободной энергии.

Таким образом, $f$--псевдофермионы в данной ситуации сами становятся коллективизированными.
В приближении среднего поля такая картина сталкивается с некоторыми  идейными трудностями.
В частности, возникает проблема исключения нефизических состояний в гильбертовом пространстве:
условие (\ref{eq:O.11}) учитывается только в среднем введением химического потенциала псевдофермионов (в рассматриваемом подходе числа псевдофермионов и электронов проводимости сохраняются по отдельности, так что $n_f=1$).
Истинная волновая функция должна получаться из среднеполевой после применения дополнительного проекционного оператора Гутцвиллера, исключающего пустые и дважды занятые $f$-состояния на узле \cite{711}.
Аналогичные проблемы картины спиновой жидкости подробно обсуждаются в книге \cite{Wen} и обзоре \cite{Wen1}, где рассмотрены флуктуационные эффекты,
которые приводят к взаимодействию с бозонными калибровочными полями.
Следует отметить, что  эффект Кондо в $s-f$ обменной модели может приводить к стабилизации экзотических состояний типа спиновой жидкости по сравнению с ситуацией в простой модели Гейзенберга \cite{711}.

В ферромагнитном состоянии имеем
\begin{eqnarray}
H-\mu \hat{n} &=&\sum_{\mathbf{k}\sigma }[(t_{\mathbf{k}}-\mu )c_{\mathbf{k}%
\sigma }^{\dagger }c_{\mathbf{k}\sigma }+W_{\sigma }f_{\mathbf{k}\sigma
}^{\dagger }f_{\mathbf{k}\sigma }  \nonumber \\
&&+V_{\sigma }(c_{\mathbf{k}\sigma }^{\dagger }f_{\mathbf{k}\sigma }+f_{%
\mathbf{k}\sigma }^{\dagger }c_{\mathbf{k}\sigma })]-\sum_{\sigma }V_{\sigma
}^{2}/I \label{eq:O.4}
\end{eqnarray}%
где  $J_{0}=J(\mathbf{{q}=0)}$,  $V_{\sigma }$  -- эффективная гибридизация,
\[
W_{\sigma }=W-\sigma J_{0}\bar{S},
\]%
$W$ -- положение $f$-уровня, отсчитываемое от химпотенциала электронов  $\mu$; величина $-W$, которая имеет порядок температуры Кондо $T_{\text{K}}$, -- химпотенциал псевдофермионов.

Соответствующая плотность состояний имеет гибридизационный вид и включает острые пики псевдофермионных состояний:
\begin{equation}
N_{\sigma }(E)=\left( 1+\frac{V_{\sigma }^{2}}{(E-W_{\sigma })^{2}}\right)
\rho \left( E-\frac{V_{\sigma }^{2}}{E-W_{\sigma }}\right)
\end{equation}%
При этом емкость зоны в два раза больше, чем у затравочной зоны.
Кроме того, функция  $N(E)$ воспроизводит и даже усиливает особенности затравочной плотности состояний.

\begin{figure}[htbp]
\includegraphics[width=3.3in, angle=0]{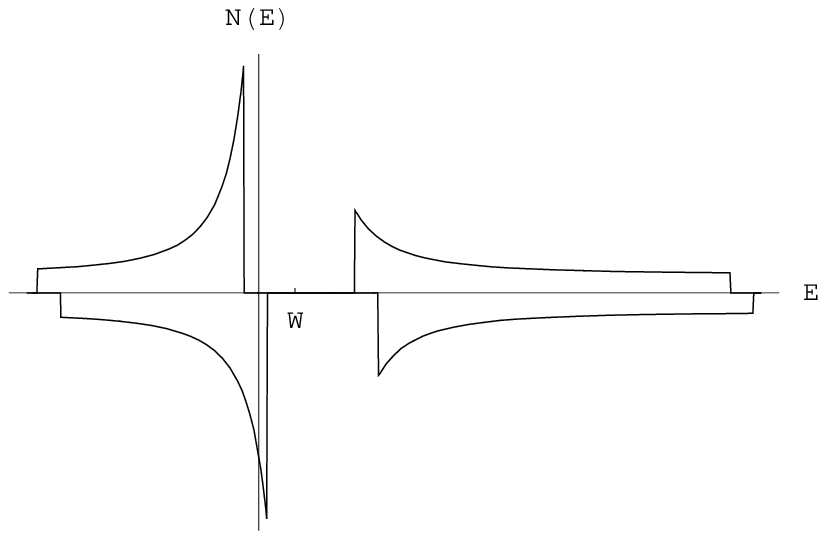}
\caption{Парциальные плотности состояний для спина вверх (верхняя половина) и спина вниз (нижняя половина) в случае прямоугольной зоны \cite{I14}. Ось ординат проведена через химический потенциал для полуметаллического ферромагнитного состояния. Отметим, что поляризации электронов проводимости  и псевдофермионов противоположны
}
\label{2}
\end{figure}

Сначала обсудим полуметаллическое ферромагнитное (ПМФ) решение, для которого химический потенциал лежит в энергетической щели для $\sigma =\uparrow $ (рис. \ref{2}), так что
\begin{equation}
W_{\downarrow }>V_{\downarrow }^{2}/(D-\mu ),~-V_{\uparrow }^{2}/\mu
<W_{\uparrow }<V_{\uparrow }^{2}/(D-\mu ),  \label{eq:O.24}
\end{equation}

Поскольку емкость гибридизационной подзоны равна единице, имеем ($n$ -- концентрация электронов проводимости)
\[
n_{\uparrow }^{f}\simeq 1-n/2,~n_{\downarrow }^{f}\simeq n/2,~\bar{S}\simeq
(1-n)/2
\]%
в то время как намагниченность электронов проводимости мала, $ n_{\uparrow }\simeq n_{\downarrow }\simeq n/2$. В этом состоянии каждый  электронов проводимости компенсирует один локализованный спин из-за отрицательного знака $s-f$ обменного параметра $ I $, а магнитное упорядочение обусловлено обменным взаимодействием некомпенсированных моментов. Такая картина напоминает ситуации в узкозонном ферромагнетике в моделях Хаббарда или $s-d$ обмена с сильным внутриатомным взаимодействием (двойной обмен). В нашем случае, однако, затравочное взаимодействие мало, но эффективное взаимодействие велико в режиме сильной связи.


Для прямоугольной полосы ($\rho (E)=1/D,~0<E<D$) условие существования ПМФ решения имеет вид ($W_{0}$ соответствует парамагнитному состоянию)
\begin{equation}
~j=J_{0}/W_{0}<\frac{2}{n(1-n)}  \label{eq:O.28}
\end{equation}
Таким образом, оно имеется при произвольно малом $J_{0}$, как и в гейзенберговском магнетике.
В приближении среднего поля полная энергия магнитного состояния всегда ниже, чем у немагнитного состояния Кондо, и падает с ростом момента:
\begin{equation}
\mathcal{E}-\mathcal{E}_{\text{non-mag}}=-J_{0}\overline{S}^{2}
\label{eq:O.29}
\end{equation}%
Энергияю ферромагнитного состояния Кондо следует также сравнивать с энергией
обычного (гейзенберговского) ферромагнитного состояния, в котором эффект Кондо подавлен ($ W = 0 $, $\bar{S}= 1/2 $). Последнее становится энергетически
выгодным при достаточно больших значениях
\begin{equation}
j>j_{c}=1/(1/4-\overline{S}^{2}).
\end{equation}%
В критической точке должен происходить магнитный фазовый переход первого рода. Последнее также подтверждается общим феноменологическим рассмотрением в работе \cite{Kirkpatrick}.

Отметим, что картина полуметаллического магнетизма тесно связана с
гибридизационным характером спектра, как и в интерметаллических $d$%
-системах \cite{RMP}.

Таким образом, ПМФ состояние достаточно устойчиво. В работе \cite{DMFT} оно было переоткрыто в модели Кондо в рамках метода DMFT и названо “spin-selective Kondo insulator”.
В работах \cite{Lacroix,Liu} оно было получено в рамках приближения Хартри-Фока, где наряду с кондовскими аномальными средними учитывались расцепления, описывающие магнитное РККИ-взаимодействие (поляризацию электронов проводимости).
Следует отметить, что в последнем подходе кондовское состояние получается только при больших $|I|$  порядка полуширины зоны, так что уже не возникает малого энергетического масштаба $T_K$. Вероятно, для построения количественной теории нужен учет перенормировки затравочных параметров, как обсуждается в разделе 3.



Теперь обсудим ненасыщенное  ферромагнитное решение с намагниченностью $\bar{S}<(1-n)/2$, где  $W_{\sigma }>V_{\sigma }^{2}/(D-\mu )$ для обеих $\sigma $
и уровень Ферми лежит в нижней гибридизационной подзоны (ниже энергетической щели), как и в немагнитном случае.

Учитывая перенормировки гибридизации в нашей модели, можно получить самосогласованное уравнение для намагниченности
\begin{eqnarray}
\frac{J_{0}\bar{S}}{W} &=&{}L(\bar{S},n),  \label{Th} \\
~L(\bar{S},n) &=&\tanh \left( \frac{1}{2\rho _{n}}\int\limits_{\mu (n+1-2%
\bar{S})}^{\mu (n+1+2\bar{S})}dE~\frac{\rho (E)-\rho}{E-\mu}%
\,\right) ,  \nonumber
\end{eqnarray}%
и выражение для переномированной температуры Кондо (энергии $f$-уровня)
\begin{eqnarray}
~W &=&W_{0}P(\bar{S},n),~ \\
P(\bar{S},n) &=&\frac{1}{2}\sum_{\sigma }\exp \left( \frac{1}{\rho }%
\int\limits_{\mu (n+1)}^{\mu (n+1+2\sigma \bar{S})}\frac{\rho (E)-\rho }{%
E-\mu }\,dE\right)   \nonumber
\end{eqnarray}%
Эти уравнения можно переписать как
\begin{equation}
~j\bar{S}=P(\bar{S},n){}L(\bar{S},n),~j=J_{0}/W_{0}.  \label{eq:O.23}
\end{equation}%
Решения с $\bar{S}\neq 0$ могут существовать, если левая и правая часть (\ref{Th}) порядка единицы, т.~е. $ J_ {0} \sim W_{0} $. Однако реальные условия для этого оказываются достаточно жесткими (ситуация похожа на приближение Хаббард-I, где имеется сильная зависимость критерия магнетизма от затравочной плотности состояний \cite{Hubbard-I:1963}). В~частности, уравнение (\ref{Th}) не имеет нетривиальных решений для $\rho (E)=\mathrm{const}$:  намагниченность возникает только за счет энергетической зависимость $\rho .$

Необходимое условие для существования решения с малыми $ \bar {S} $ имеет вид
\begin{equation}
k_{n}=\left. \frac{dL(\bar{S},n)}{d\bar{S}}\right\vert _{\bar{S}=0}=\frac{1}{%
\rho _{n+1}\rho _{n}}\frac{\rho _{n+1}-\rho _{n}}{\mu _{n+1}-\mu _{n}}>0
\label{K}
\end{equation}%
где $\rho _{n}=\rho (\mu _{n})$.
Таким образом, по сравнению с критерием полуметаллического ферромагнетизма, который определяется  глобальным поведением $\rho (E)$, критерий ферромагнетизма с малыми $\bar {S} $ определяется $\rho (E\simeq \mu _{n+1})$ (что соответствует "<большой"> поверхности Ферми). Ситуация  для
возникновения решений с малыми $ \bar {S} $ более благоприятна при больших $ n $  и в присутствии высоких узких пиков $\rho (E)$.

\begin{figure}[htbp]
\includegraphics[width=3.3in, angle=0]{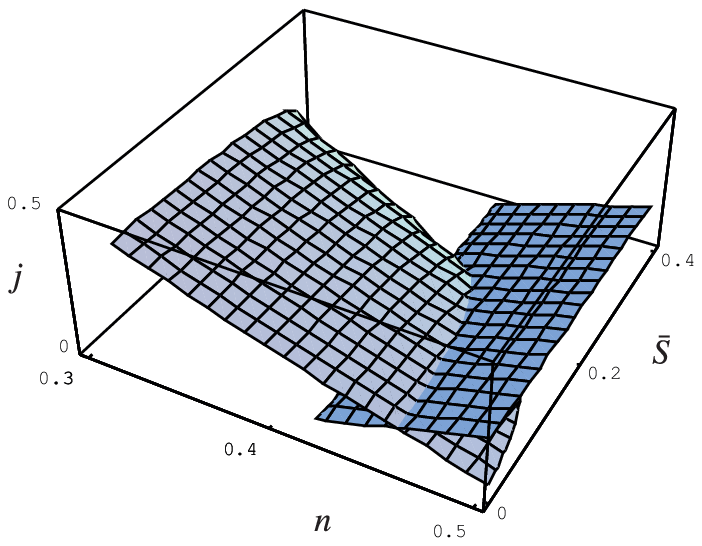}
\caption{ Диаграмма поверхности существования ненасыщенного ферромагнитного решения для затравочной полуэллиптической зоны согласно (\protect\ref{eq:O.23}) (левый лист) \cite{I14}. Насыщенное (полуметаллическое) решение существует выше правого листа}
\label{6}
\end{figure}

Для простых решеток с симметричной затравочной плотностью состояний ненасыщенное решение может существовать только в области  $ n <1/2 $ (когда $\rho _{n+1}<\rho _{n}$). В частности, для полуэллиптической зоны при  $J_{0}\rightarrow 0$ ферромагнетизм исчезает для $n>0.5$, а насыщенное ферромагнитное решение возникает для $n<0.42$ (рис. \ref{6}).

Условие для ферромагнетизма с $\bar{S}\rightarrow 0$ имеет вид $k=j$,
так что ферромагнитное решение начинается с $j=k$, а с уменьшением $ j $ момент растет.
Такое необычное поведение связано со своеобразным характером ненасыщенного решения. В этом состоянии имеются две конкурирующие тенденции. Рост обменного расщепления $ 2J_0 \bar{S} $ с~$j$ толкает химический потенциал в энергетической щели, но острый псевдофермионный пик не позволяет пересечь себя и остается выше уровня Ферми. Таким образом, $ \bar{S}$ должно уменьшаться с ростом~$j$.
Отметим, что в случае РККИ обмена (когда $ J \sim I^2 \rho $, а $T_K$ экспоненциально мала) рост $ j $ соответствует уменьшению $|I|$, и поэтому картина не столь парадоксальна, как кажется на первый взгляд.



\begin{figure}[htbp]
\includegraphics[width=3.3in, angle=0]{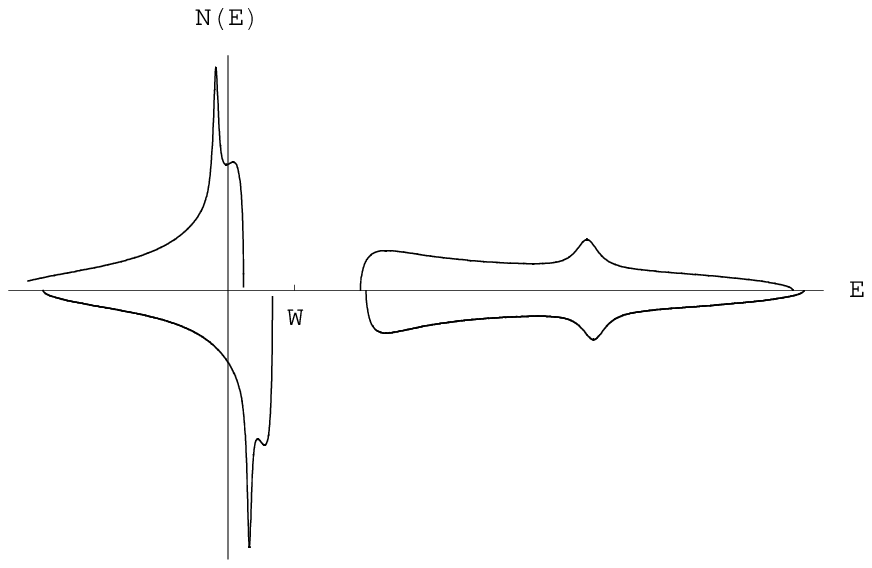}
\caption{Парциальные плотности состояний для полуэллиптической зоны с лоренцевским пиком
}
\label{4}
\end{figure}

Как показывают численные расчеты \cite{I14}, для простой квадратной и кубической решеток ненасыщенное решение существует лишь внутри области существования насыщенного и проигрывает ему по энергии.
Для возникновения магнетизма с малым моментом благоприятна ситуация, когда имеется узкий пик плотности состояний на $\mu _{n+1}$ или чуть выше, так что упорядочение отодвигает его с уровня Ферми, который оказывается в слабом локальном минимуме для состояний со спином вверх (рис. \ref{4}).

\begin{figure}[htbp]
\includegraphics[width=3.3in, angle=0]{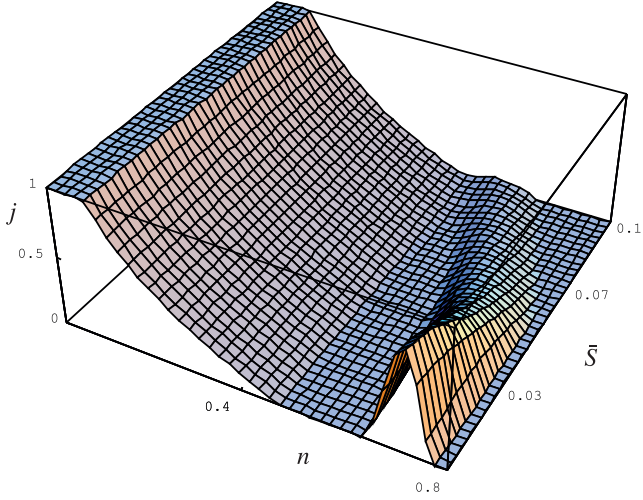}
\caption{ Диаграмма поверхности существования ненасыщенного ферромагнитного решения для полуэллиптической зоны с лоренцевским пиком ($\rho (E)=\rho _{0}(E)+ h\Gamma/[(E-E_{1})^{2}+\Gamma ^2]$) при  $\Gamma=0.01D,~E_{1}=0.8D,~h=0.01$ \cite{I14}}\label{9}
\end{figure}

Область ненасыщенного решения для затравочной плотности состояний с лоренцевским пиком показана на рис. \ref{9}. Видно, что магнетизм с малым моментом возникает в довольно широкой области вблизи значений $n$, где $\mu _{n+1}=E_{1}$, причем состояние ПМФ в этой области концентраций не существует, по крайней мере, при малых $j$.

Таким образом, кондовский ферромагнетизм имеет ряд особенностей по сравнению с картиной Стонера для обычных систем коллективизированных электронов. В частности, зависимость критерия ферромагнетизма от затравочной плотности состояний гораздо сложнее. Причина состоит в том, что решающую роль для этого критерия играет зависимость эффективной гибридизации от спина.

Отметим,  что рассмотренный подход дает универсальное описание в терминах отношения $j \sim J_{0}/T_{K}$ и функций, которые зависят только от затравочной плотности состояний, но не от $s-f$ обменного параметра  $I$.


Гибридизационная форма электронного спектра (наличие пиков плотности состояний) в кондо-решетках подтверждается многочисленными экспериментальными исследованиями: прямыми оптическими данными \cite{Mar}, наблюдением больших электронных масс в измерениях эффекта де Гааза-ван Альфена и т.д. Иногда трудно четко различить режимы промежуточной валентности (ПВ) и Кондо в формировании магнетизм, так как $ f$-состояния в обоих случаях играют важную роль в электронной структуре
вблизи уровня Ферми. Наличие реальной гибридизации между $ s, d $- и $ f $ -состояниям в ПВ системах может привести к пику в затравочной плотности состояний, который, в свою очередь, будет влиять на псевдофермионный ферромагнетизм. Возможный пример -- ферромагнитная система  CeRh$_{3}$B$_{2}$ с малым моментом и высокой температурой Кюри, которая связывалась с сильной $d-f$ гибридизацией.
В то же время анализ нейтронных данных для нее не обнаружил намагниченности на узлах родия и бора, так что ферромагнетизм связан с упорядочением локальных моментов церия, а не является коллективизированным магнетизмом в зоне родия, как предполагалось ранее \cite{Alonso}.

\begin{figure}[htbp]
\includegraphics[width=3.3in, angle=0]{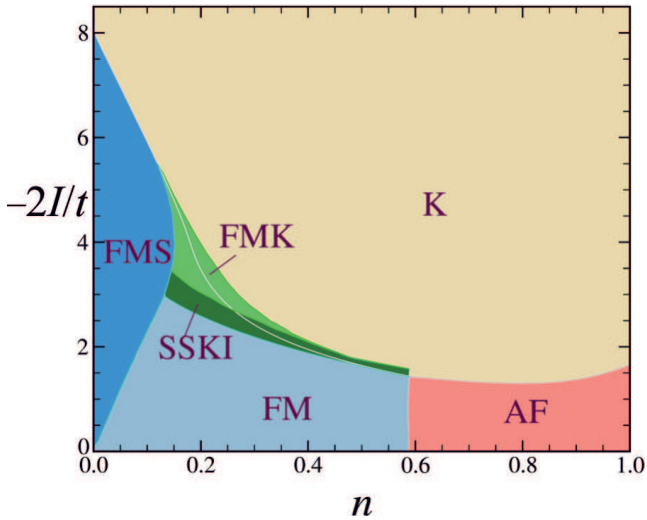}
\caption{
Магнитная фазовая диаграмма для модели Кондо на квадратной решетке \cite{Lacroix}.
FMS и FM -- насыщенный и ненасыщенный ферромагнетизм без кондовской компенсации,
SSKI и FMK -- фазы сосуществования эффекта Кондо и насыщенного (полуметаллического) и ненасыщенного ферромагнетизма соответственно
}
\label{lacroix}
\end{figure}

В случае антиферромагнитного упорядочения электронный спектр искажен как гибридизационной (кондовской), так и антиферромагнитной щелью.
Здесь аналитическое исследование уравнений среднего поля затруднено, и они требуют численного решения.
В~отличие от ферромагнетика, для  антиферромагнетизма с малыми моментами  можно пренебречь зависимостью гибридизации от $\sigma $, так как поправки вследствие спиновой поляризации
имеют структуру $(J_{\mathbf{Q}} \bar{S})^2/
(t_{\mathbf{k}+\mathbf{Q}}-t_{\mathbf{k}})$ $\mathbf{Q}$ -- волновой вектор магнитной структуры) и пропорциональны $(J_{\mathbf{Q}}\bar{S})^2/W$. Таким образом, критерий
антиферромагнетизма принимает обычную форму:
\begin{equation}
J_{\mathbf{Q}}\chi _{\mathbf{Q}}>1,
 \label{eq:6.105}
\end{equation}
где $\chi _{\mathbf{Q}}$~"---  восприимчивость невзаимодействующих $f$"~псевдофермионов в эффективной гибридизационной модели.
Благодаря гибридизационным пикам вклад от межзонных переходов
оказывается большим:
$\chi _{\mathbf{Q}}\sim 1/T_{\text{K}}$. Следовательно, антиферромагнетизм появляется при
\begin{equation}
J_{\mathbf{Q}}>\nu T_{\text{K}},
 \label{eq:6.107}
\end{equation}
где постоянная $\nu $ порядка единицы определяется зонной структурой.
Следует, однако, отметить, что согласно численным результатам для квадратной решетки \cite{Lacroix} момент не является малым, а фазовая диаграмма достаточно сложна (рис. \ref{lacroix}), причем сосуществование кондовского состояния и антиферромагнетизма имеет место при достаточно больших $|I|$ и $J$.
Для последовательного рассмотрения здесь важен также учет спиральных магнитных состояний \cite{Igoshev}. Эти проблемы требует дальнейшего исследования.

Нетривиальной проблемой является вопрос, как влияют на магнетизм флуктуаций за пределами приближения среднего поля.
Простые спин-волновые поправки, которые обсуждались в работе \cite{608}, дают лишь формально небольшие вклады  в намагниченность основного состояния порядка $j \ln j$ (которые к тому же отсутствуют в ПМФ состоянии). Более последовательное рассмотрение флуктуаций может быть выполнено с использованием подхода вспомогательных бозонов и $ 1 / N $-разложения в рамках периодических моделей Андерсона или Коблина-Шриффера \cite{Coleman2}.
Для картины экзотической ферми-жидкости флуктуации изучались в работе \cite{Sachdev} в рамках псевдофермионнного представления с одновременным использованием вспомогательного бозона (типа поля Хиггса).

Поведение кондовских магнетиков при конечных температурах обычно трактуется в рамках теории типа Стонера с гибридизационным спектром \cite{Lacroix,Liu}, что по большому счету  не оправдано: флуктуационные эффекты должны быть даже важнее, чем в основном состоянии (как и в обычных зонных магнетиках \cite{26}). Они обсуждаются в \cite{Sachdev}, где показано, что температурный фазовый переход из кондовского состояния в обычную ферми-жидкость на самом деле является плавным (кроссовером).

\section{Заключение}

Магнетизм решеток Кондо представляет собой сложный феномен: они проявляют черты магнетиков как с коллективизированными электронами, так и с локализованными моментами. При определенных условиях магнитная неустойчивость может иметь место при очень малых  (даже по сравнению с $ T_ {K} $) значениях межузельного обмена $ J $, что характерно для гейзенберговских систем. С другой стороны, магнитное упорядочение весьма чувствительно к электронной структуре, как и для коллективизированных систем.
Особенно ярко эта двойственность проявляется в подходах, использующих представление псевдофермионов, которые практически становятся реальными и принимают участие в формировании поверхности Ферми.

Описанные механизмы формирования магнитного состояния с подавленным моментом насыщения $\mu_s$
значительно отличаются от обычного механизма для зонных  ферромагнетиков, которые, как предполагается, в грубом приближении описываются теорией  Стонера.
Однако, поскольку и энергетический спектр новых фермиевских
квазичастиц, и эффективное взаимодействие между ними претерпевают
сильные перенормировки, для кондовских ферромагнетиков неприменим критерий Стонера (или, в случае антиферромагнетизма, критерий Оверхаузера для формирования волны спиновой плотности) с затравочными параметрами. Скорее его можно применять к эффективной модели в сильно коррелированном состоянии с "<большой"> поверхностью Ферми, однако и здесь ситуация оказывается существенно более сложной.

Как мы видели, современная теория описывает магнитные решеток Кондо как сильно коррелированные системы.  Так как существует непрерывный переход между кондо-решетками и "<стандартными"> системами коллективизированных электронов (в частности, обычные
паулиевские парамагнетики могут рассматриваться как системы с
высоким $T_{\text{K}}$~"--- порядка энергии Ферми), возникает
вопрос относительно роли, которую многоэлектронные эффекты играют
в "<классических"> слабых зонных магнетиках, подобных ZrZn${}_2$.
Может оказаться, что близость основного состояния к точке
стонеровской неустойчивости, т.~е. малость $\mu_s$, в
последних системах возникает из-за перенормировок параметра
взаимодействия $U$ и химпотенциала, а не из-за случайных затравочных
значений $N(E_{\text{F}})$. Действительно, в такую случайность
трудно поверить, поскольку отклонение от граничного условия
Стонера $UN(E_{\text{F}})=1$ является крайне малым.

Все это дает и новый взгляд на физику обычных слабых коллективизированных магнетиков: их можно рассматривать не~с зонной точки зрения, а с точки зрения локальных магнитных моментов, которые почти скомпенсированы.
Сейчас уже стало обычным совместно обсуждать решетки Кондо
и зонные магнетики \cite{Ohkawa,Vojta} и трактовать UPt${}_3$, CeSi${}_x$ и
CeRh${}_3$B${}_2$ как слабые коллективизированные магнетики (см.,
напр., \cite{613}). Таким образом, теория кондовских магнетиков приобретает важное значение для общей теории магнетизма.

Автор благодарен М.И. Кацнельсону и В.Н. Никифорову -- соавторам работ по физике решеток Кондо -- за многочисленные обсуждения.
Работа выполнена в рамках государственного задания ФАНО (тема «Квант», No. 01201463332) и частично поддержана грантом УрО РАН 15-8-2-9.

\end{document}